\documentclass[preprint,12pt,3p]{elsarticle}

\usepackage{amssymb}
\usepackage{multirow}
\usepackage{booktabs}
\usepackage[justification=centering]{caption}
\usepackage{siunitx}
\usepackage{listings}
\usepackage{color}
\usepackage{array}
\usepackage{hyperref}

\newcolumntype{C}[1]{>{\centering\let\newline\\\arraybackslash\hspace{0pt}}m{#1}}

\definecolor{dkgreen}{rgb}{0,0.6,0}
\definecolor{gray}{rgb}{0.5,0.5,0.5}
\definecolor{mauve}{rgb}{0.58,0,0.82}

\journal{Journal of Systems and Software}

\begin{document}

\begin{frontmatter}

\title{Exploring Software Reusability Metrics with Q\&A Forum Data}

\author[label1]{Matthew T. Patrick}
\address[label1]{Department of Dermatology, University of Michigan, Ann Arbor MI}
\ead{mattpat@umich.edu}

\begin{abstract}
Question and answer (Q\&A) forums contain valuable information regarding software reuse, but they can be challenging to analyse due to their unstructured free text. Here we introduce a new approach (LANLAN), using word embeddings and machine learning, to harness information available in StackOverflow. Specifically, we consider two different kinds of user communication describing difficulties encountered in software reuse: `problem reports' point to potential defects, while `support requests' ask for clarification on software usage. Word embeddings were trained on 1.6 billion tokens from StackOverflow and applied to identify which Q\&A forum messages (from two large open source projects: Eclipse and Bioconductor) correspond to problem reports or support requests. LANLAN achieved an area under the receiver operator curve (AUROC) of over 0.9; it can be used to explore the relationship between software reusability metrics and difficulties encountered by users, as well as predict the number of difficulties users will face in the future. Q\&A forum data can help improve understanding of software reuse, and may be harnessed as an additional resource to evaluate software reusability metrics.
\end{abstract}


\end{frontmatter}

% \linenumbers

\section{Introduction}
Software reuse is an important strategy for decreasing development costs and increasing productivity, as well as avoiding defects and improving software quality \cite{mohagheghi_2007}. It was originally envisaged as a way to make software development more efficient through modular components that can be used over and over again in mass production \cite{mcilroy_1969}, rather than rewriting functionality that already exists, as was (and is) common practice. Nevertheless, there is a cost to software reuse, as it is necessary to develop and maintain `glue code' that connects the reusable component with the software under development \cite{svahnberg_2016}. There is also a concern that software written by other people may contain unknown bugs, such that it is difficult to ensure the quality of applications constructed from reused components.

The potential for bringing existing software components and knowledge to a new project depends on their `reusability' \cite{frakes_2005}. Various reusability metrics have been suggested \cite{ampatzoglou_2018}, based on factors ranging from the software's complexity (structural code quality, dependencies, size etc.) through to its understandability (interface complexity and documentation). Previous researchers \cite{lemley_1997} have considered software reuse in terms of direct costs (integrating/adapting existing software components in the new application, versus rewriting them from scratch) and indirect costs (the potential for errors and bugs in reused versus newly developed software). We hypothesise the direct costs of software reuse are likely to depend on its understandability (i.e. the software interface), while the indirect costs may be associated with its complexity (under the assumption that more complex software is more likely to go wrong).

To investigate our hypothesis, we introduce a new approach (LANLAN: Lexical ANalysis for LAbelling iNquiries) that extracts information from question and answer (Q\&A) forums. LANLAN classifies questions into `problem reports' (indicating possible defects) and `support requests' (asking for help in understanding how to use the software). Software that has a lot of support requests demonstrates direct costs, since users/reusers have difficulty applying it, while software that has many problem reports may be more likely to harbor bugs (i.e. indirect costs). By applying statistical techniques to test the association between Q\&A messages and features derived from static analysis relating to complexity and understandabilty, we hope to be able to explore the relationship between problem reports / support requests and software reuse.

In early research, data about problems experienced during software development and reuse was expensive or difficult to obtain, being primarily extracted from corporate testing activities \cite{endres_1975} or classified military records \cite{goel_1979}. By contrast, the rise of open source software has made data publicly available for mining \cite{antoniol_2004}: version control repositories (such as GitHub\footnote{GitHub: \url{https://github.com/}}) contain information about changes made and the reasons for making them, whilst bug tracking databases (e.g. Bugzilla\footnote{Bugzilla: \url{https://www.bugzilla.org}}) record observed failures along with attempts to identify and address their cause. Researchers have applied various metrics (lines of code, coupling, churn etc. \cite{radjenovic_2013}) to analyse this data, and machine learning algorithms (e.g. SVM or Random Forest \cite{bowes_2017}) have been used in an attempt to improve software quality (and hence reusability).

Techniques which aim to improve software quality include those which direct developers towards specific parts of their code more likely to contain defects \cite{bowes_2017}\cite{hall_2012} or model the overall quality and health of a software project \cite{jansen_2014}\cite{bedoya_2014}, but evaluation of these techniques depends on the quality and size of available data. Bug report and version control repositories are often affected by various biases \cite{nguyen_2010}. For example, experienced developers are more likely to submit bug reports, whilst novice users often feel discouraged to contribute for fear of condescension \cite{lotufo_2012}. Bug reports can sometimes contain contradictory claims or be impossible to reproduce \cite{schugerl_2008}\cite{sun_2011}. For example, in one study, 40\% of files marked as defective in five open source projects never actually contained a bug \cite{herzig_2013}. Q\&A forums have their own biases and accuracy issues, since they depend on how users express their questions. However, by combining multiple sources of data together, we should be able to improve the robustness of our analyses when evaluating effective metrics for software reusability.

Community-driven resources, such as mailing lists and Q\&A forums, allow users to describe problems and work together to fix them \cite{abdalkareem_2017}. Issues are frequently described within these resources without being reported in any other database. For example, Bachmann et al. \cite{bachmann_2010} observed 16\% of defects in the Apache web server were addressed in the software's mailing list instead of its bug tracking system. Q\&A forums also contain information about software developer/user communities and their interaction \cite{vasilescu_2013}, which might be helpful for understanding the social dynamics of software reuse. However, it can be difficult to derive meaningful categorisations from the unstructured text in social media, due to subtle nuances of communication and natural language \cite{wang_2019}. In this paper, we propose a new approach (LANLAN) to mine information directly from existing Q\&A forums and classify posts automatically using statistical and machine learning techniques from the field of natural language processing.

We evaluate LANLAN on two large open source projects (Eclipse and Bioconductor) through cross-validation and testing on different software from which the model was trained. We apply novel approaches (association analysis and growth curve modelling) to interpret the results and find key differences between the occurrence of problem reports and support requests that may be useful in improving the reusabilty of software. The remainder of this paper is organised as follows: Section 2 introduces the background and related work, Section 3 explains our approach, Section 4 describes our evaluation procedures, Section 5 provides the results and discussion, Section 6 explores the threats to validity, Section 7 presents our conclusions, and Section 8 lists the code availability.

\section{Background and Related Work}
Q\&A forum mining has frequently been applied to analyze user behaviour, from early research into Usenet \cite{whittaker_1998}, through to more recent investigations of contributor motivations \cite{treude_2011}, collective knowledge \cite{anderson_2012} and the effectiveness of code examples \cite{nasehi_2012} in StackOverflow. Machine learning techniques have also been applied to make predictions from this data. For example, Yang et al. \cite{yang_2011} applied various classifiers to predict which questions will remain unanswered, whereas Zhang et al. \cite{zhang_2015b} used Latent Dirichlet Allocation (a topic modelling approach) to predict duplicate questions. Q\&A forum mining has been used to assist software developers in an IDE prompter for Java \cite{ponzanelli_2014} and an interactive programming tool for Python \cite{rong_2016}. In common with these studies, we apply machine learning techniques to Q\&A forum data. However, as far as we are aware, our paper represents the first attempt to use data mined from Q\&A forums to predict difficulties faced during software reuse.

LANLAN extracts useful information by combining Q\&A forum data with other features, e.g. the GitHub repository. GitHub is often used in repository mining, due to its large size and accessibility through an open API \cite{kagdi_2007}. For example, Ray et al. \cite{ray_2014} used GitHub to explore the relationship between programming language and code quality, and Zanetti et al. \cite{zanetti_2013} applied network analysis and machine learning to predict the quality of bug reports. Zhang et al. \cite{zhang_2015} used topic models to predict the interest and experience of developers as related to specific bug reports, assigning the most appropriate developer to fix a particular bug. In software ecosystem research \cite{manikas_2016}\cite{mens_2014}, software projects are compared to natural ecosystems, modelling their development using techniques normally applied in ecology or evolutionary theory. We also adapt techniques typically used in the natural sciences (growth modelling and association analysis) to interpret the data we have collected.

Zeller \cite{zeller_2013} discussed the challenges involved in mining software repositories further. For example, it can often be difficult to distinguish fixes from other changes, such as those that add new features or refactor the code. Linking repositories to a bug database can help identify which changes relate to bugs, but even when bug databases are used, a large proportion of fixes are not recorded in them. For the Eclipse project, less than half of fixes could be linked to an entry in the bug database \cite{bird_2009}. Zeller \cite{zeller_2013} argues software repository mining is useful despite these issues, but it should be augmented by seeking input from project insiders or using approaches such as keyword matching (to predict bug fixes from other changes). We augment repository mining with information from Q\&A forums and show our machine learning approach to be more effective than simple keyword matching.

Central to our approach are numerical representations of words, known as embeddings \cite{goth_2016}, that take inspiration from ordinary language philosophy \cite{wittgenstein_1953} and structuralist linguistics \cite{firth_1957}. Word embeddings capture the semantics of words from a corpus according to their context (i.e. the words that surround them) \cite{goth_2016}. Information is distributed among a small (fixed) number of weights, with the assignment of values to these weights providing a distinct vector (and therefore semantics) for each word. A key advantage of word embeddings (compared to other natural language processing techniques, such as named entity recognition, or sentence parsing) is that they provide a uniform representation, which can easily be used to train advanced machine learning models (e.g. for sentiment analysis \cite{giatsoglou_2017}).

The earliest word embedding approaches used global factorization. For example, Latent Semantic Analysis (LSA) \cite{deerwester_1990} constructs a matrix of counts for the number of times words occur in each document, then applies Singular Value Decomposition (SVD) to factorize it into vectors for each word. Global factorization is a coarse-grained approach for modeling semantics, and is especially limited if the documents being analyzed are large. More detailed information can be obtained through the analysis of local context (i.e. words that occur near each other), for example the skip-gram approach \cite{mikolov_2013}, which was previously applied to documentation from the Java Development Kit for code retrieval \cite{nguyen_2017}. However, there is a danger predicting the context of one word at a time will miss information available through global statistics. We aim to find a middle ground between these two strategies using Global Vectors for Word Representation (GloVe) \cite{pennington_2014} to incorporate data at both the global and local scale. To the best of our knowledge, this paper represents the first time GloVe has been applied to the field of software engineering.

Global Vectors for Word Representation (GloVe) \cite{pennington_2014} has been used on tasks as diverse as annotating videos from free text descriptions \cite{hendricks_2017}, to identifying implicit human bias/stereotyping \cite{greenwald_2017}. It takes advantage of local information (by counting word co-occurrences in their local context) as well as global (i.e. aggregated) statistics. By contrast with the skipgram technique, which predicts words from their context one at a time, GloVe uses a highly parallelizable matrix factorization approach. However, instead of factorizing a global document-word count matrix (as with LSA), GloVe factorizes a matrix of word-word co-occurrences ($X_{ij}$), produced using a sliding window.

LANLAN identifies features that may be indicative of difficulties in software reuse, because they are associated with support requests or problem reports. Opinion differs as to the effectiveness of using features of the software to improve quality (i.e. static analysis): Rahman et al. \cite{rahman_2014} suggested static analysis can complement statistical defect prediction, whereas Johnson et al. \cite{johnson_2013} suggested it is underused in practice, due to problems with false positives. One recent work in this area \cite{ray_2016} likened the characteristics of program code to natural language, and suggested entropy measures may be used to predict software defects. We also utilise techniques from natural language processing in our research, but apply them to Q\&A forum messages rather than the code itself. Although we apply LANLAN to analyzing software reusability, it also has potential for the development process as a whole.

\section{Our Approach}
\begin{figure*}[ht]
\caption{Data Flow Diagram for LANLAN}
\centering
\includegraphics[width=\linewidth]{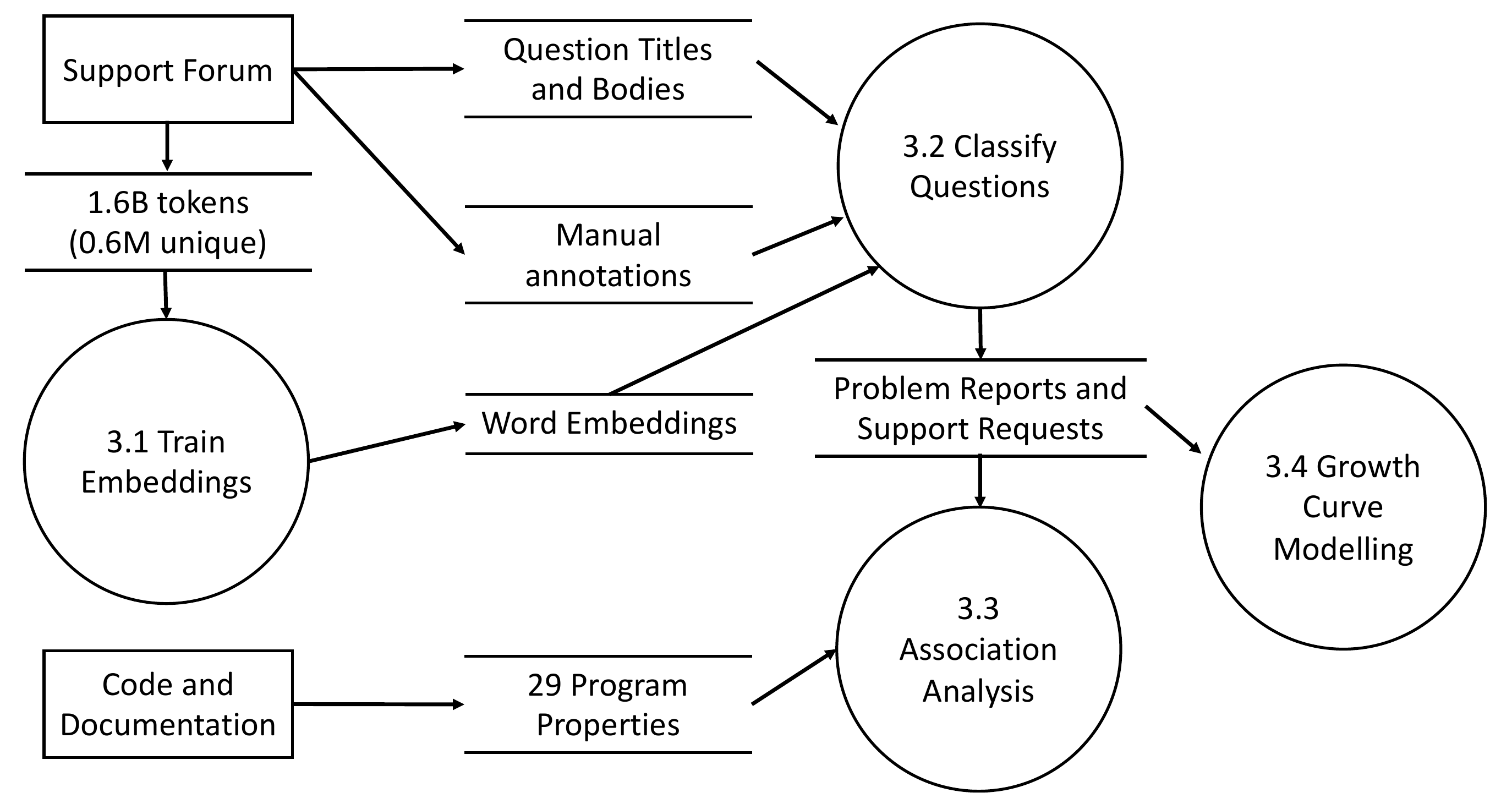}
\label{workflow}
\end{figure*}

Figure \ref{workflow} shows the data flow for our approach introduced in this paper for classifying Q\&A forum posts (LANLAN). In particular, we aim to distinguish questions that indicate a potential defect in the software (e.g. ``I think there may be a bug in...") from those asking for help in achieving their goals for reuse (e.g. ``please could you tell me how to..."). We call the first category \textbf{problem reports} and the second category \textbf{support requests}. Word embeddings are trained using a large corpus of text from Q\&A forum messages. We then pre-process the questions asked about each program, mapping their words to the corresponding embedding, and creating features for each question. Machine learning is performed to produce a prediction model, then the results are analysed using growth curves and association analysis.

\subsection{Training Word Embeddings}
We trained word embeddings on questions submitted to StackOverflow, consisting of 1.6 billion tokens, with a vocabulary of 0.6 million unique words. StackOverflow questions were downloaded from the Stack Exchange Data Dump\footnote{\url{https://archive.org/details/stackexchange}} and then parsed using HTMLParser in Python to remove the XML tags. Prior to training, each word was tokenized and transformed into lower case. We then removed all characters not in the roman alphabet or specific punctuation marks (full stops, question marks or exclamation marks). All numbers (regardless of length) were replaced by the token `0' (so as to avoid creating a separate embedding for each individual number, and to treat the presence of any number as the feature we wish to encode) and code blocks were replaced by the token `$<$code$>$'. We also transformed all types of exception and error (e.g. \emph{NullPointerException}) into the words `exception' and `error', to ensure LANLAN can easily be transferred to other datasets (which may use different exception and error types).

\begin{figure*}[ht]
\caption{GloVe Matrix Factorization}
\centering
\includegraphics[width=.85\linewidth]{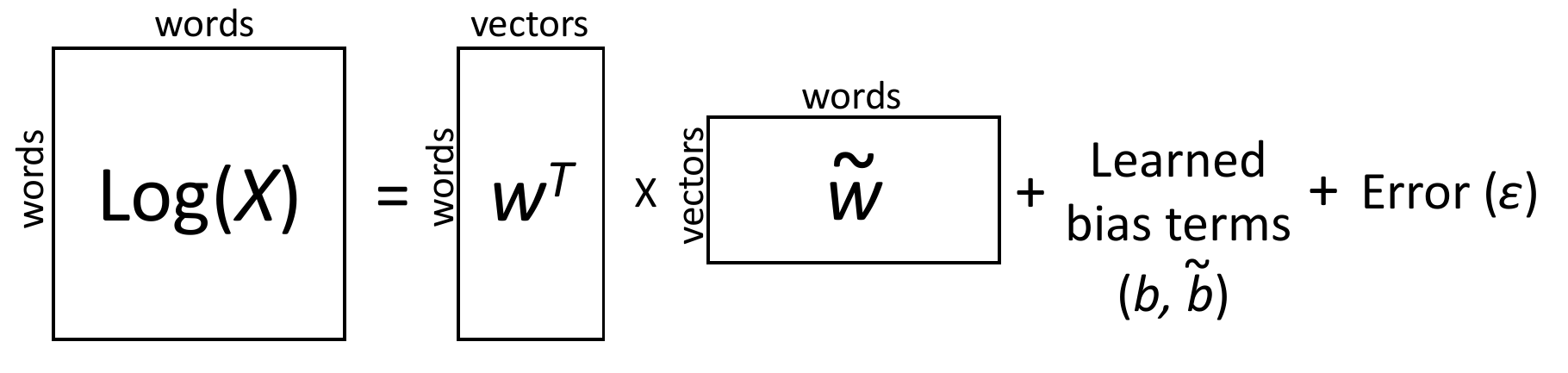}
\label{factorization}
\end{figure*}

GloVe generates two sets of word embeddings ($w$ and $\tilde{w}$) as a result of the matrix factorization (see Figure \ref{factorization}). The embeddings are optimized by learning bias terms ($b_i$ and $\tilde{b}_j$) for each set, such that the difference between the log of the original word-word matrix ($X$) and the matrix reconstructed from the embeddings and bias terms is as small as possible, i.e. the error term ($\epsilon$) is reduced. This approach can be represented as an optimization function (Equation \ref{eq:glove}), and once the word embedding sets ($w_i$ and $\tilde{w}_j$) are optimized, they are added together to improve their accuracy. Furthermore, a weighting function [$f(x)=(x/x_{max})^\alpha$ if $x<x_{max}$, 1 otherwise] is applied to avoid learning only from common word pairs (where $x_{max}=100$ and $\alpha=\frac{3}{4}$); for more information see Pennington et al. \cite{pennington_2014}. In our experiments, we trained word embeddings as 200-dimensional vectors and used the default window size of 15 words, because these settings were found to be effective in previous research \cite{pennington_2014}.
\begin{equation}
    \sum_{i,j=1}^{V}f(X_{ij})(w_i^T\tilde{w}_j+b_i+\tilde{b}_j-\log(X_{ij}))^2=\epsilon
    \label{eq:glove}
\end{equation}

\subsection{Classifying Q\&A Forum Posts using Word Embeddings}
Our aim is to use the information contained within word embeddings to classify forum posts into support requests and problem reports. Questions are pre-processed in the same way as the training dataset, except code, numeric and punctuation tokens are removed (these tokens are used only for context in training and not to evaluate question semantics). Each question consists of a title and a body: we produced a set of features for these components by calculating the mean embedding from the words they contain. This approach has previously been found to be effective at comparing the similarities between short texts (similar to our Q\&A forum questions) \cite{kenter_2015}. Unlike taking the sum of values in word embeddings, the mean is not influenced by the length of the text. Since each word embedding consists of a vector of 200 numerical values, this gives us 400 features for each question.

LANLAN uses these features to distinguish questions asking for clarification on software usage (support requests) from those referring to potential defects in the software (problem reports), by applying a variety of classification algorithms through a machine learning framework in R (MLR \cite{bischl_2016}). Each algorithm is evaluated through stratified 10-fold cross-validation: dividing the questions at random into ten equal-sized partitions, then validating a model on each partition (one at a time) after training it on the remaining data. Stratification ensures the same proportion of class labels are included in each (randomly selected) partition, which is particularly important when class labels are imbalanced (i.e. most questions posted to Q\&A forums do not indicate defects). We train LANLAN on manually annotated questions and, to evaluate whether our approach may be transferred to other programs and datasets, we also train classification models on one program and then test them on others.

\subsection{Association Analysis}
Association analysis is a technique for identifying properties significantly correlated with a particular trait. For example, in bioinformatics it helps discover which genetic markers affect the observable characteristics of an organism \cite{balding_2006}. In our work, we are interested in finding program properties (potential software reusability metrics) which could lead to an increase or decrease in the number of support requests and problem reports. To achieve this, we fit a linear model to the data and test whether the regression coefficients for each property are equal to zero (using a t-statistic). The results can then be used to infer the probability each property is significantly correlated with the number of questions that report potential defects (problem reports) or ask for help using the software (support requests).

Linear models assume each data point is independent (we ensured this by treating each thread in the Q\&A forum as an individual sample); the residuals (i.e. difference between the fitted model and the data) should follow a normal distribution (we tested this using a Q-Q plot); the variance for the residuals should be homogeneous and there should be a linear relationship between the dependent and independent variables (this was tested using a plot of residuals against fitted values). It is important to ensure these assumptions are met for us to have confidence in our evaluation of the significance of each property.

We use Bonferroni correction \cite{dunn_1961} to address the multiple comparisons problem (i.e. the more properties we test, the more likely p-values will be significant by chance). This involves dividing the standard significance threshold (0.05) by the number of comparisons (i.e. properties) to identify those which have a high likelihood of being significant. This is a conservative measure, since some program properties are likely to be correlated with each other. As well as applying association analysis to each property, we also identify subsets of properties that are almost as descriptive of the underlying factors as the entire set. We do this by evaluating the multiple $r^2$ value of all sets of properties of size five. This procedure is applied separately for problem reports and support requests to identify the most important properties for understanding the factors behind the number of questions in each category.

\subsection{Growth Curve Modelling}
To illustrate how the classification models produced by LANLAN may be used to predict the rate at which support requests and problem reports occur, we analyse the resulting data using growth curve models. Growth curve modelling offers a way to understand and compare the dynamics of problem reports and support requests over the software's lifetime. Although this technique has rarely been applied in software engineering, it is popular in a variety of fields, such as economics, public health, ecology and social demography \cite{panik_2014}.

\begin{figure}[ht]
\caption{Example of Growth Curve Modelling}
\vspace{-6mm}
\centering
\includegraphics[width=.7\linewidth]{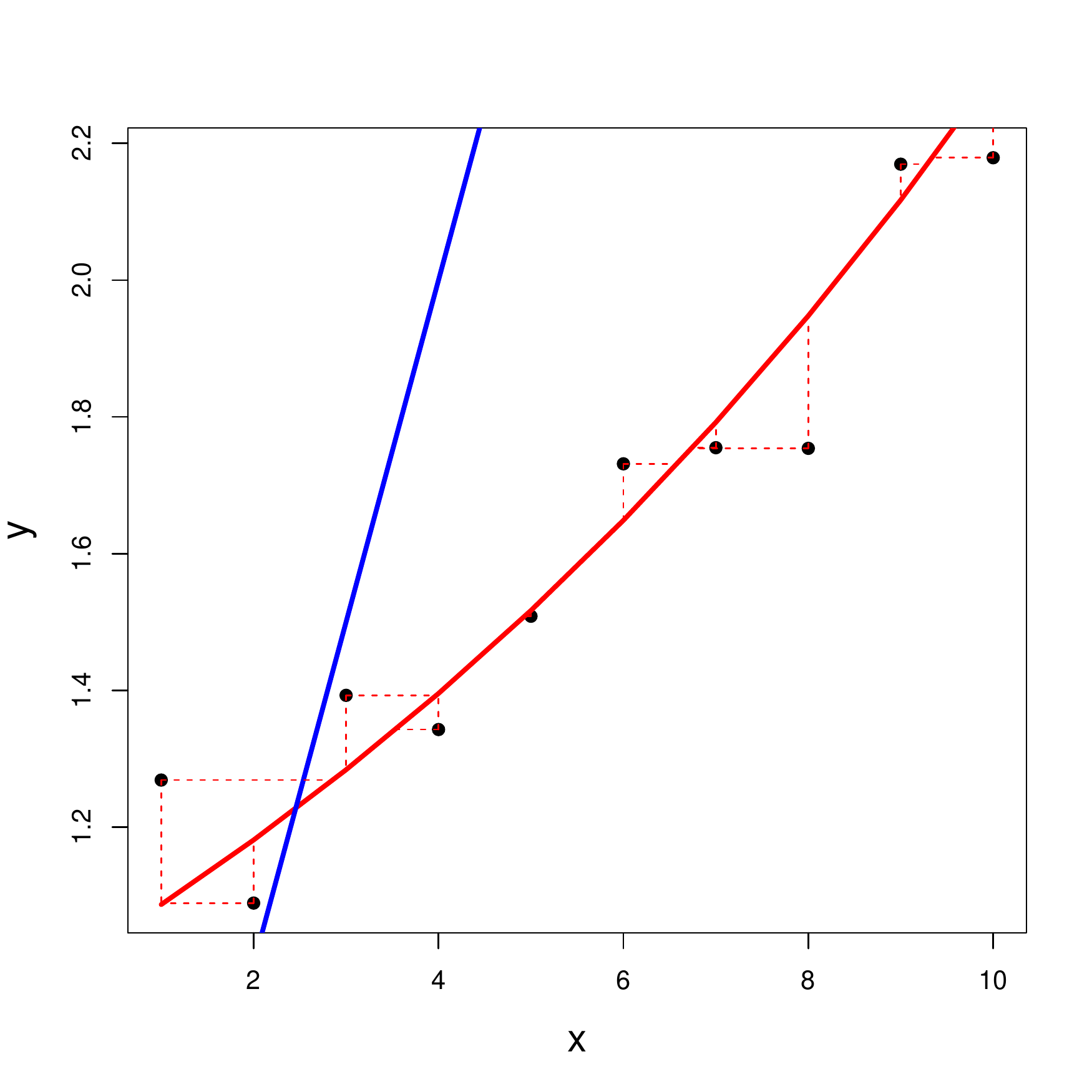}
\label{fittingExample}
\end{figure}

Growth curve modelling attempts to identify the curve that most closely fits the data, by optimizing over a number of parameters. One way to achieve this is through least squares estimation, i.e. minimizing the sum of the square distances between each data point and the curve. Figure \ref{fittingExample} plots two different curves against a series of data points (in black). The red curve is an instance of an exponential model, while the blue curve is from a linear model. Although some of the points lie closer to the linear (blue) model, if we were to take the sum of squared distances across x and y (dashed lines), we would find the objective value to be smaller for the exponential (red) model. Hence, this particular exponential model is a better fit for the data. Using this approach, we can identify not only the most appropriate family of curves, but also their optimal parameters. We model the cumulative number of problem reports and support requests for each program by fitting a generalised logistic growth model (Equation \ref{eq:generalised_logistic}). The generalised logistic growth model is highly flexible and by changing its parameter values ($\kappa$, $\delta$ and $\beta$), can represent many different forms of growth \cite{panik_2014}.

\begin{equation}
    Y(t)=\frac{\kappa}{[1-\exp{(-\beta(t-\delta))}]^{-5}}
    \label{eq:generalised_logistic}
\end{equation}

Where possible we fit the growth curves through non-linear least squares estimation (using the stats package in R). The success of this technique depends, at least in part, on the starting values chosen for each parameter. We used the maximum number of problem reports / support requests (for each program) as a suitable starting value for $\kappa$, because it represents the y-asymptote (i.e. the number the curve will tend towards as time increases); we chose two different growth rate values for $\beta$ (0.05 and 0.01) and set the starting value for $\delta$ to 0 ($\delta$ is considered the delay parameter). The delay parameter was set at zero as our initial assumption is that the first date recorded for each software is the date its usage started to grow, and the two values for the growth rate were chosen to explore the range of possible rates at which software usage grows (some software will be adopted quickly, whereas others take longer to become popular).

Whenever non-linear least squares estimation failed to converge, we applied Bayesian parameter estimation (using the R interface to the JAGS MCMC library\footnote{rjags: \url{https://CRAN.R-project.org/package=rjags}}). Rather than just fitting a curve to the data points available for each package, MCMC takes into account prior knowledge about the distribution of each parameter. We set the prior distributions according to distributions of fitted parameter values from nls (lognormal for $\kappa$, normal for $\delta$ and gamma for $\beta$). The mean of the resulting parameter values from 100 chains (i.e. executions) of MCMC were calculated and used for each program.

\section{Evaluation}
\subsection{Worked Examples}
To evaluate the robustness of LANLAN, we selected two large open source projects (Eclipse and Bioconductor) as worked examples. Eclipse has previously been used as the subject of studies into software reuse \cite{hummel_2008}\cite{martinez_2016}\cite{martinez_2017} and Ye et al. \cite{ye_2014} created a database of bug reports (BugDB) for Eclipse, which now forms part of the NASA Promise repository\footnote{NASA Promise Repository: \url{http://openscience.us/repo}}. The programs in this project include AspectJ (an aspect-oriented programming extension), Birt (a business intelligence and reporting tool), JDT (a suite of Java development tools), SWT (a widget toolkit) and Tomcat (a web application server).

Our second worked example, Bioconductor \cite{gentleman_2005}, consists of a large collection of (over 1,400) bioinformatics and molecular biology software packages, written for different purposes and by different people. However, Bioconductor provides a standard interface from which a wide range of statistics may be derived. This offers us the opportunity to evaluate the factors that affect the number of problem reports and support requests.

One way to divide open source software reuse \cite{brown_2002} is by whether it is a pre-planned strategy of a popular commercial product (e.g. the Eclipse suite) or the ad-hoc process of finding software developed to perform a specific task (such as the individual software packages that make up Bioconductor). By including two projects (Eclipse and Bioconductor) that are very different from each other, we should have a better idea of whether LANLAN will work on a wide range of software projects.

\subsection{Research Questions}
\begin{enumerate}
\item[\textbf{RQ1:}]
{\bf How accurate is LANLAN at classifying forum posts?}
Before we can have confidence in our technique (LANLAN), we need to make sure it accurately identifies questions which refer to defects (problem reports) as opposed to those asking for clarification on software usage (support requests). We achieve this by evaluating LANLAN on four different programs (two from the Eclipse suite and two from Bioconductor). To determine whether LANLAN can reliably identify problem reports from support requests, we compare the predicted categorisations against manually annotated labels. This research question evaluates the steps described in Section 3.2.

\item[\textbf{RQ2:}]
{\bf Can support requests and problem reports be used to evaluate software reusability metrics?}
The number of support requests and problem reports for each software package may help to provide insight into the impact of different features on reusability. We apply association analysis to a variety of program properties, to see which correlate with each kind of post. We also look for groups of properties that together are almost as representative as the entire set. By analysing problem reports and support requests in this way, we should be able to identify program properties that are important to consider when attempting to improve software reusability. This research question evaluates the steps described in Section 3.3.

\item[\textbf{RQ3:}]
{\bf Can LANLAN predict how support requests and problem reports will grow in the future?}
Given the previous numbers of Q\&A forum questions, it would be useful to predict how many will occur in the future, since this could guide developers where to focus their efforts to improve reusability. For example, if problem reports are predicted to grow faster than support requests, it may be more efficient to spend time now identifying and fixing bugs, whereas if the situation is reversed, it would be better to focus effort on improving documentation or simplifying the software interfaces. We evaluate the accuracy of these models by using them to make predictions based on partial data (i.e. up to a certain point in time) and then comparing the results with the actual number of problem reports and support requests subsequently observed. This research question evaluates the steps described in Section 3.4.

\item[\textbf{RQ4:}]
{\bf How useful is the distinction of support requests and problem reports to potential reusers?}
Not only is it necessary to evaluate the accuracy with which problem reports / support requests can be identified and predicted, but it is important to consider whether doing so provides useful information for software reuse. Adapting software to new purposes can be challenging and time-consuming, so any added expense of applying LANLAN has to be worthwhile for the benefits it provides. To investigate this, we sampled 10 support requests and 10 problem reports at random from AspectJ to explore in more detail. We consider the relationship these Q\&A messages may have to reuse activities and ask what this means for the success rate or difficulty of reusing code with many problem reports or support requests.
\end{enumerate}

\subsection{Experimental Setup}
Word embeddings were trained with a vector size of 200, using 593,767 StackOverflow questions from the 31 August 2017 data dump. We then extracted all questions related to each program in our study individually, with the StackOverflow and Bioconductor API. In total, we categorised 45,093 questions for Eclipse and 23,556 questions for Bioconductor (as problem reports or support requests). Of this data, we manually annotated all the questions related to 4 programs: AspectJ, EclipseJDT, edgeR and PROcess; this represents 4,630 questions (or 7\% of the total). Each program had its questions manually annotated three times (on separate occasions by the author), and then the annotations were combined by consensus (see Table \ref{tab:annotations}). To predict whether the remaining questions were related to defects, we benchmarked 24 different classification algorithms using cross-validation in MLR. Subsequently, 29 program properties were evaluated for their association with problem reports and support requests, using three covariates to control confounding factors (See Section 5.2).

\begin{table}[ht]
\centering
\begin{tabular}{c|cc|c}
\toprule
& \textbf{Problem} & \textbf{Support} & \textbf{Total} \\
& \textbf{Reports} & \textbf{Requests} & \textbf{Questions} \\\midrule
AspectJ & 602 & 1,879 & 2,481 \\
EclipseJDT & 97 & 637 & 734 \\
edgeR & 97 & 683 & 780 \\
PROcess & 105 & 530 & 635 \\\midrule
\textbf{Total} & \textbf{901} & \textbf{3,729} & \textbf{4,630}
\end{tabular}
\caption{Numbers of problem reports and support requests annotated manually}
\label{tab:annotations}
\end{table}

\section{Results and Discussion}
\subsection{Accuracy of Our Approach (Answer to RQ1)}
\begin{figure}[ht]
\caption{Benchmarking 24 classifiers on AspectJ}
\centering
\includegraphics[width=.7\linewidth]{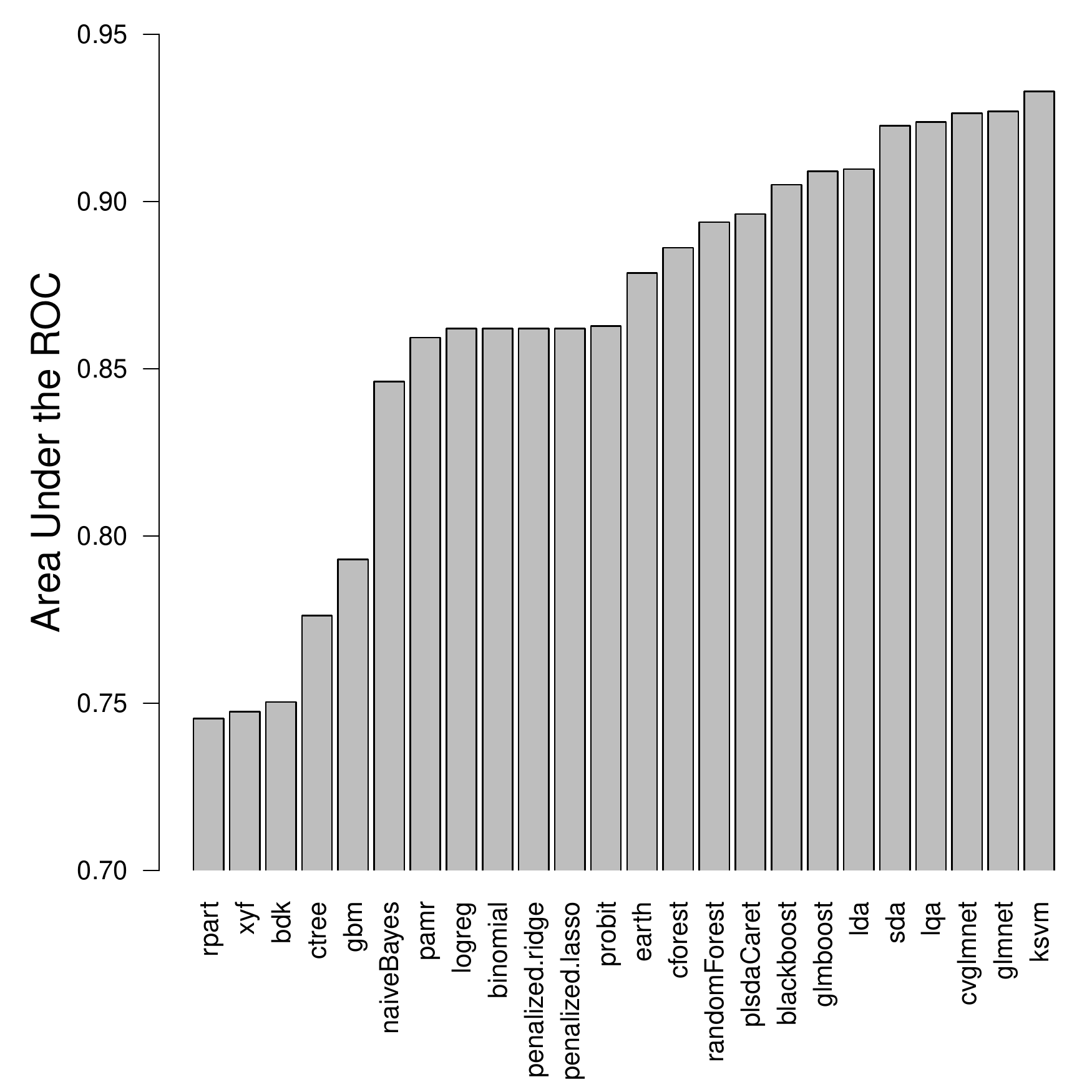}
\label{fig:benchmark}
\end{figure}

LANLAN automatically predicts whether questions submitted to Q\&A forums are related to defects in the software (problem reports) or if they request more general help/advice (support requests). To evaluate its accuracy, and answer RQ1, we compared these predictions against the consensus annotations described in Section 4.3. First, we trained our prediction model on AspectJ (the annotated program which has the largest number of user-submitted questions). In our benchmark results (see Figure \ref{fig:benchmark}), 19 out of the 24 classification algorithms (80\%) achieved an Area Under the Receiver Operator Curve (AUROC) above 0.8 in 10-fold cross validation, highlighting the robustness of LANLAN. In particular, the Support Vector Machine (SVM) had the highest AUROC (0.930).

An alternative strategy \cite{zeller_2013}, previously suggested for distinguishing repository commits which correct faults from those that make other changes, looks for indicator keywords in the text (specifically `bug', `problem', or `fix'). We compared LANLAN with two different versions of this alternative: in the first version (keyword matching), we counted posted questions as problem reports if any of these keywords were present; whilst in the second version (keyword features), we created a new prediction model using the number of times keywords occur as features. As with our approach, we extracted features from the title and body separately, and then combined them together in the prediction model. It is not possible to compare the AUROC of the keyword matching approach (since it does not provide class probabilities for each feature), but the maximum AUROC achieved in a benchmark on the keyword features approach was only 0.562 (using Random Forest in 10-fold cross-validation). This suggests our approach (LANLAN) to be substantially more effective than the alternative.

We also evaluated precision (the proportion of questions correctly identified as problem reports) and recall (the proportion of problem reports correctly identified as such). Together, these two metrics represent the ability of an approach to classify questions accurately and find the majority of problem reports. LANLAN provides considerably higher precision and recall than both the alternative approaches. In particular, 81\% of questions LANLAN indicates to be problem reports are correct (precision), as compared to 71\% for the keyword features model and only 42\% for keyword matching. Furthermore, LANLAN identifies 72\% of problem reports correctly (recall), as opposed to 31\% for keyword matching and only 9.6\% for the keyword features model. It is interesting the keyword feature model performs better on precision, whereas keyword matching achieves higher recall. This could be because keyword matching includes all questions that contain the specified keywords, whereas the keyword feature model is trained in a more sophisticated way. Crucially, LANLAN outperforms both the alternatives on precision and recall.

To evaluate how well LANLAN generalises to other programs, we applied the model trained on AspectJ to the Eclipse Java Development Tools (JDT) user interface and a mathematical program from Bioconductor (edgeR) for differential expression analysis (see Table \ref{tab:machine_learning_results}). This reduced the AUROC slightly (from 0.930 for AspectJ to 0.889 for JDT and 0.921 for edgeR), but larger differences were seen in precision and recall. On JDT, precision was 0.659 (compared to 0.810 for AspectJ); precision was unaffected for edgeR, but recall fell to 0.330 (compared to 0.720 for AspectJ). These differences are likely to be caused by variations in the language used to communicate problems for each program. For example, in JDT a `Quick Fix' is a pop-up that helps users with their code, so features that indicate the request for a fix (problem report) in AspectJ may point to support requests in JDT. Bioinformatics software is used primarily by scientists rather than software engineers, so different vocabulary can often be used to describe problem reports, thus reducing the recall.

\begin{figure}[hb]
\vspace{5mm}
\begin{minipage}{.5\textwidth}
\centering
\caption{ROC curve for first three programs}
\includegraphics[width=\linewidth]{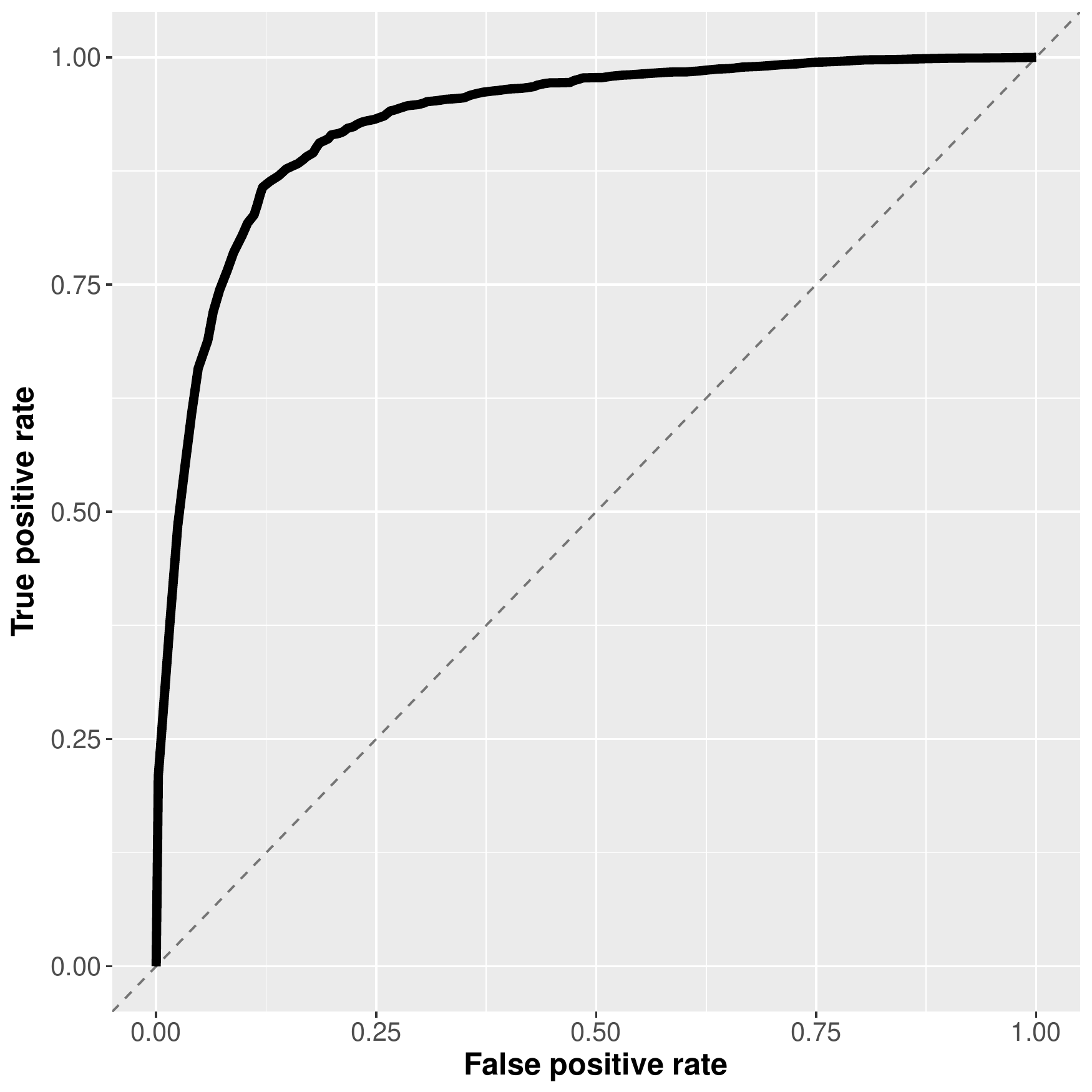}
\label{fig:roc}
\end{minipage}%
\begin{minipage}{.5\textwidth}
\centering
\caption{Precision/Recall for PROcess}
\includegraphics[width=\linewidth]{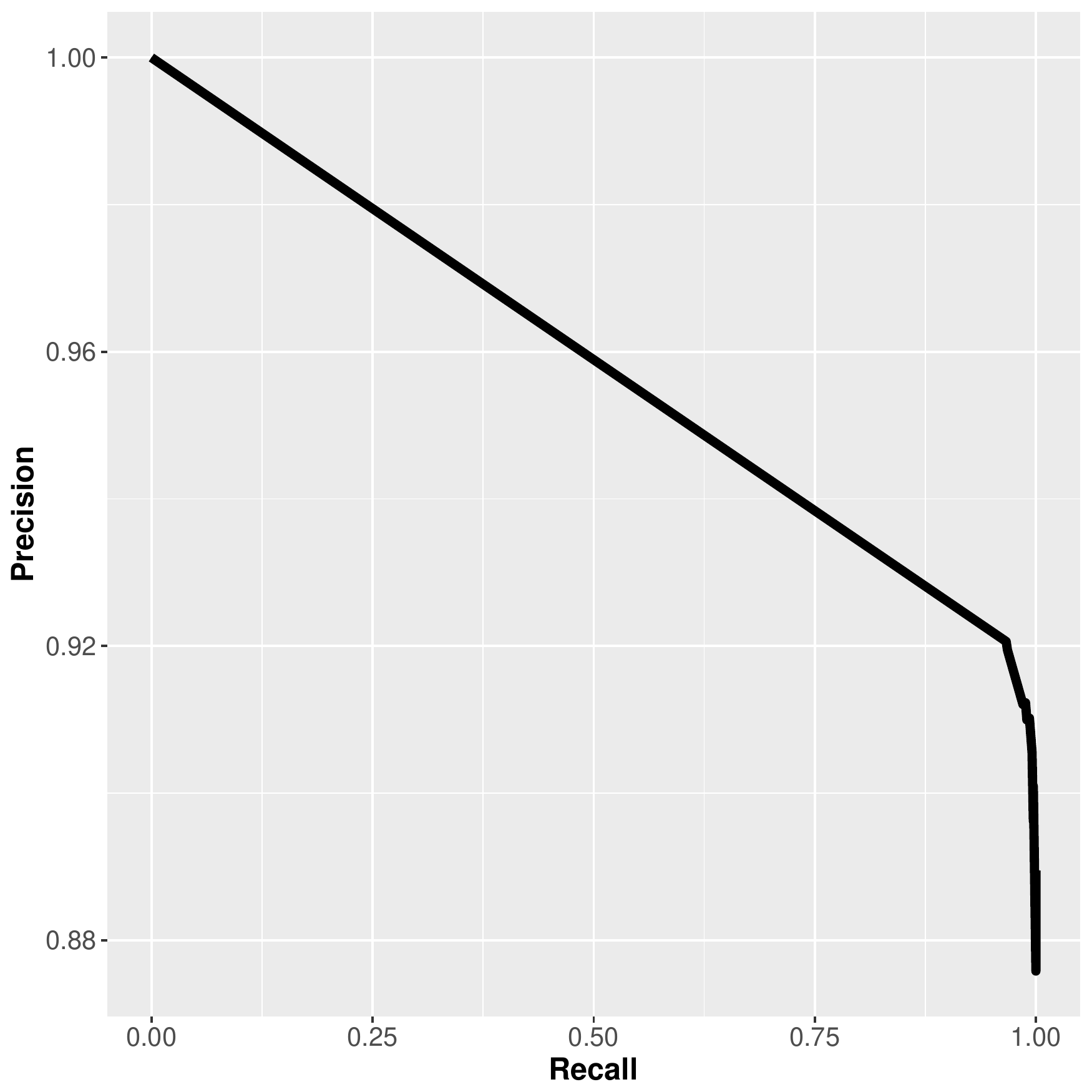}
\label{fig:precision_recall}
\end{minipage}
\end{figure}

We strengthened our approach by training the prediction model on a range of different programs. When trained on AspectJ, JDT and edgeR, then tested on a different Bioconductor program (PROcess), the AUROC, precision and recall were higher than for any individual program before. Figure \ref{fig:roc} shows the combined ROC curve when training on AspectJ, JDT and edgeR; the precision/recall curve when testing this model on PROcess can be seen in Figure \ref{fig:precision_recall}. Following this approach, our prediction model is accurate on the programs for which it has been trained, as well as being robust when applied to new programs.

\begin{table}[ht]
\vspace{2mm}\centering
\begin{tabular}{cc|c|c|c|c}\toprule
& & \textbf{AspectJ$^*$}&\textbf{JDT}&\textbf{edgeR}&\textbf{PROcess$^{**}$}\\\midrule
\multirow{3}{*}{LANLAN} & \textbf{AUROC} & 0.930 & 0.889 & 0.921 & 0.970\\
& \textbf{Precision} & 0.810 &  0.659 & 0.842 & 0.919\\
& \textbf{Recall} & 0.720 &  0.577 & 0.330 & 0.752\\\midrule
\multirow{3}{*}{Keyword features} & \textbf{AUROC} & 0.562 & 0.684 & 0.680 & 0.692\\
& \textbf{Precision} & 0.714 &  0.630 & 0.750 & 0.857\\
& \textbf{Recall} & 0.096 &  0.351 & 0.247 & 0.309\\\midrule
\multirow{3}{*}{Keyword matching} & \textbf{AUROC} & \emph{NA} & \emph{NA} & \emph{NA} & \emph{NA}\\
& \textbf{Precision} & 0.423 &  0.338 & 0.329 & 0.324\\
& \textbf{Recall} & 0.312 &  0.505 & 0.560 & 0.657\\
\end{tabular}
\caption{Evaluating how well our machine learning approach (LANLAN) generalises to new programs\newline($^*$trained and tested on the same program; $^{**}$trained on the previous 3 programs)}
\label{tab:machine_learning_results}
\end{table}

\subsection{Identifying Features with the Most Impact on Reusability (Answer to RQ2)}
We applied the results of our models to evaluate the impact of 32 features (from each software package) on the number of problem reports and support requests in Bioconductor. The features were chosen to reflect previously proposed software reusability metrics, while allowing for the finer details of each metric to be explored with respect to R. For example, Chidamber and Kemerer \cite{chidamber_1994} proposed evaluating the complexity of software according to the number of methods per class, as well as communication and inheritance between classes. Since R is not an object-oriented language, we instead explore the number of lines of code per file, as well as the dependencies (both mandatory and suggested) between software packages. Following the suggestion of Buse and Weimer \cite{buse_2010}, we include measurements of code churn as a surrogate for readability, but we also consider other features, such as comments and whitespace, as well as vignettes (separate documentation, illustrating examples of use).

Various metrics (number of files, functions, blank lines, comments and lines of code) are recorded separately for the R and compiled code, then static analysis reports are generated (using Codetools\footnote{Codetools: \url{https://CRAN.R-project.org/package=codetools}} and Goodpractice\footnote{GoodPractice: \url{https://github.com/mangothecat/goodpractice}}) from the R code, as well as the (mean, maximum and total) cyclomatic complexity. We collected test coverage from CodeCov\footnote{CodeCov: \url{https://codecov.io/}} and used the GitHub API\footnote{GitHub API: \url{https://developer.github.com/v3/}} to count the number of months the package has been active, its downloads and unique downloads (by IP address), as well as additions and deletions (churn).

Whilst some features are extracted directly from the data we collected (e.g. the total number of comments in the R code of a particular package), other features are combined from multiple data (e.g. the number of comments per line of code). These features were extracted to give us information about the rates and proportions of certain properties of a package, rather than just their absolute value. Features were extracted using a variety of Linux tools, such as grep, sed and awk.

\begin{table}
\centering
\begin{tabular}{c|cc|cc}\toprule
    \textit{Package}& \multicolumn{2}{c|}{\textbf{Problem Reports}} & \multicolumn{2}{c}{\textbf{Support Requests}} \\
    \textit{Features}&P-value&Effect Size&P-value&Effect Size\\\midrule
Churn.Adds&0.031&2.15&0.016&1.80\\
Churn.Dels&0.024&2.27&0.012&1.87\\
Churn.Adds.per.Week&0.010&2.59&0.002&2.30\\
Churn.Dels.per.Week&0.017&2.40&0.005&2.10\\
R.Files&\textbf{$<$.001}&4.77&\textbf{$<$.001}&4.85\\
R.Blanks&\textbf{$<$.001}&4.41&\textbf{$<$.001}&3.82\\
R.Comments&0.028&2.21&\textbf{$<$.001}&2.74\\
R.LOC&\textbf{$<$.001}&5.59&\textbf{$<$.001}&5.28\\
R.LOC.per.File&0.369&0.90&0.499&0.50\\
R.Blanks.per.LOC&0.298&-1.04&0.335&-0.72\\
R.Comments.per.LOC&0.682&-0.41&0.494&0.53\\
Compiled.Files&0.247&-1.16&0.113&-1.22\\
Compiled.Blanks&0.459&-0.74&0.199&-0.99\\
Compiled.Comments&0.361&-0.91&0.163&-1.07\\
Compiled.LOC&0.649&-0.46&0.400&-0.65\\
Compiled.LOC.per.File&0.348&-0.95&0.500&-0.5\\
Compiled.Blanks.per.LOC&0.874&0.16&0.565&-0.43\\
Compiled.Comments.per.LOC&0.670&0.43&0.938&0.06\\
Max.Cyclomatic.Complexity&0.392&0.86&0.063&1.39\\
Mean.Cyclomatic.Complexity&0.307&-1.02&0.532&-0.47\\
Count.Cyclomatic.Complexity&0.066&1.84&0.003&2.28\\
Total.Cyclomatic.Complexity&\textbf{$<$.001}&3.87&\textbf{$<$.001}&3.93\\
Codetools.Problems&\textbf{$<$.001}&3.91&\textbf{$<$.001}&2.46\\
Test.Coverage&0.642&-0.47&0.234&0.93\\
Goodpractice.Problems&0.235&1.19&0.171&1.02\\
Vignettes&\textbf{$<$.001}&4.84&\textbf{$<$.001}&4.75\\
Depends&0.026&2.23&0.022&1.74\\
Imports&\textbf{$<$.001}&6.78&\textbf{$<$.001}&5.52\\
Suggests&\textbf{$<$.001}&3.54&\textbf{$<$.001}&3.62\\
\bottomrule\end{tabular}
\caption{Association analysis results comparing the number of problem reports and support requests with various program properties (potential reusability metrics)}
\label{tab:association_results}
\end{table}

Association analysis was applied to each of the 32 features (see Table \ref{tab:association_results}), to identify those with the most affect on problem reports and support requests (scaled using log10 transform, to ensure a linear relationship). The number of months active, downloads and IP addresses were used as covariates, to find significant features independent of the amount of usage of each package. Following Bonferroni correction, 8 features were identified as significant for problem reports and support requests (R.Files, R.Blanks, R.LOC, Total.Cyclomatic.Complexity, Codetools.Problems, Vignettes, Imports and Suggests). In addition, R.Comments was significant for support requests. R.Files counts the number of files of R code in the software, R.LOC does the same for the number of lines of code, R.Blanks for the number of blank lines and R.Comments for the number of comments. Cyclomatic complexity \cite{mccabe_1976} is a long established measure of code complexity, based on the number of independent paths through the program, whereas Codetools is a modern software package for identifying potential problems in R code. Vignettes are documentation files in R, giving specific examples of usage, while Imports count the number of other packages the software requires to work and Suggests counts the number of packages that are suggested (but not required) by the software. Overall, these results imply that problem reports and support requests are both affected by complexity, but documentation has a greater impact on support requests. 

Interestingly, test coverage and churn had no significant association with problem reports and support requests, despite the intuitive link between testing and quality, and the prevalence of churn in fault prediction \cite{hall_2012}. We repeated our analysis, using only packages that have $>0\%$ test coverage, but the p-values did not improve. One potential reason for this is that CodeCov uses a simplistic coverage criteria (statement coverage), that may not be sufficient to test the software thoroughly. Also, scientific software is particularly difficult to test, because of challenges in constructing a suitable test oracle \cite{ammann_2016}.

\begin{figure}[ht]
\vspace{2mm}
\caption{Distribution of Top 10 Subsets (by Multiple $R^2$) of Length Five}
\centering
\includegraphics[width=\linewidth]{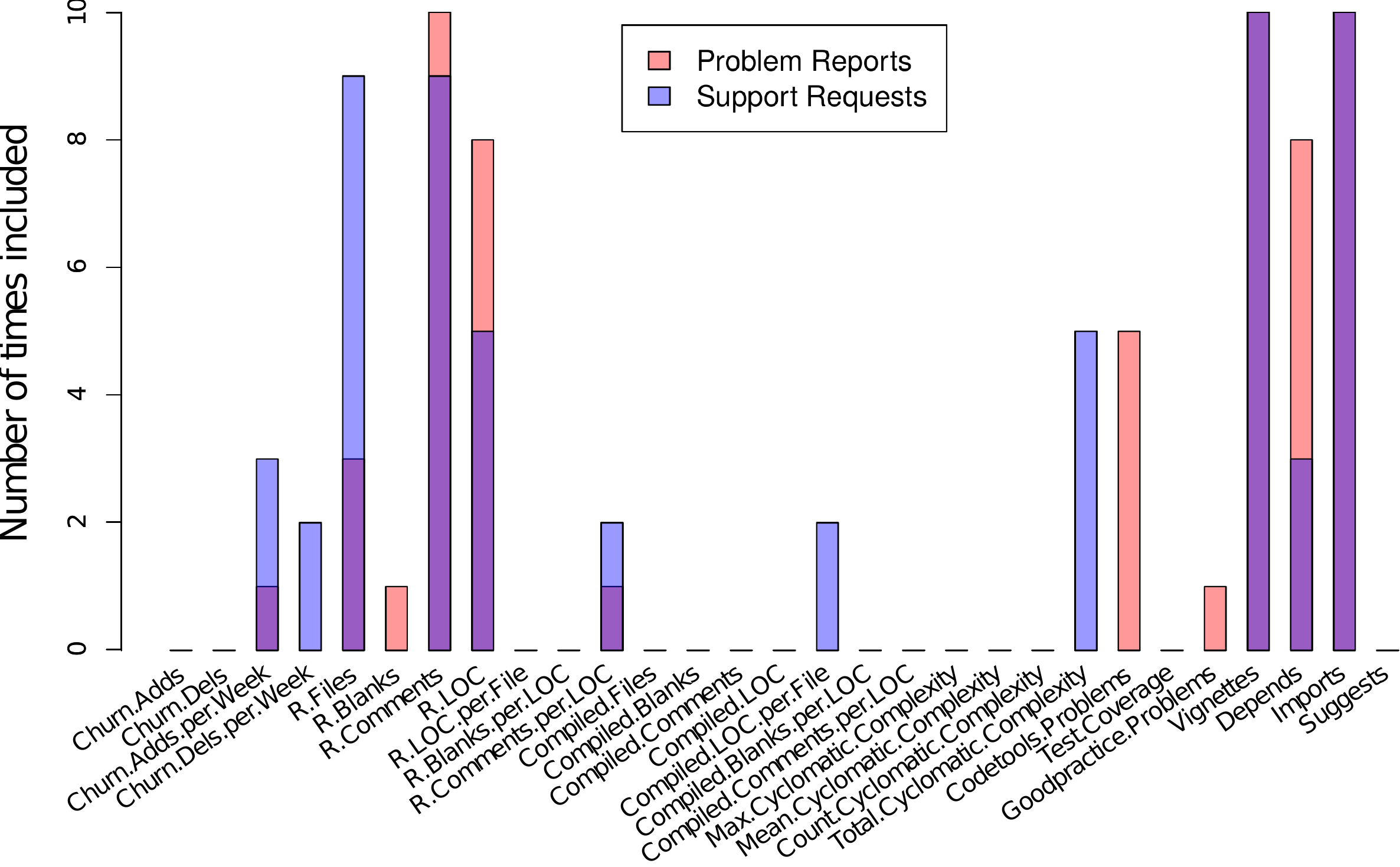}
\label{top10five}
\end{figure}

In addition to applying association analysis to each feature independently, we also evaluated subsets of features together. Figure \ref{top10five} shows the number of times each feature was included in the top 10 subsets (by multiple $R^2$ value) out of all subsets with 5 features (118,755 in total). Interestingly, Codetools.Problems appears in half of the top 10 subsets for problem reports, but none for support requests. By contrast, Total.Cyclomatic.Complexity appears in half of the subsets for support requests, but none for problem reports. When software is more complicated (has higher cyclomatic complexity) it is generally more difficult to understand and requires more explanation (through support requests), but that does not necessarily mean it contains more defects (if it is developed well). By contrast, the Codetools library highlights poor coding style, which is more likely to indicate (and possibly cause) defects. This finding goes against our initial hypothesis that problem reports would be more closely associated with complexity, while support requests would be associated with understandability, and indicates the importance of avoiding over-complicated program structure when developing software for understandability and reusability.

Selecting the top subset of features for problem reports (R.Comments, R.LOC, Codetools.Problems, Vignettes and Imports) and support requests (R.Files, R.Comments, Total.Cyclomatic.Complexity, Vignettes and Imports) reduces the training data to 17\% (from 29 features down to 5). However, 96\% and 97\% of the $R^2$ value was maintained, for problem reports and support requests respectively. This indicates these features are a good representation of the underlying factors behind the number of problem reports and support requests, and hence are likely to be important for making predictions about reusability.

\subsection{Modelling Support Requests and Problem Reports (Answer to RQ3)}
\begin{figure}[ht]
\caption{Example Growth Curve Fitting (on ArrayExpress)}
\centering
\includegraphics[width=.6\linewidth]{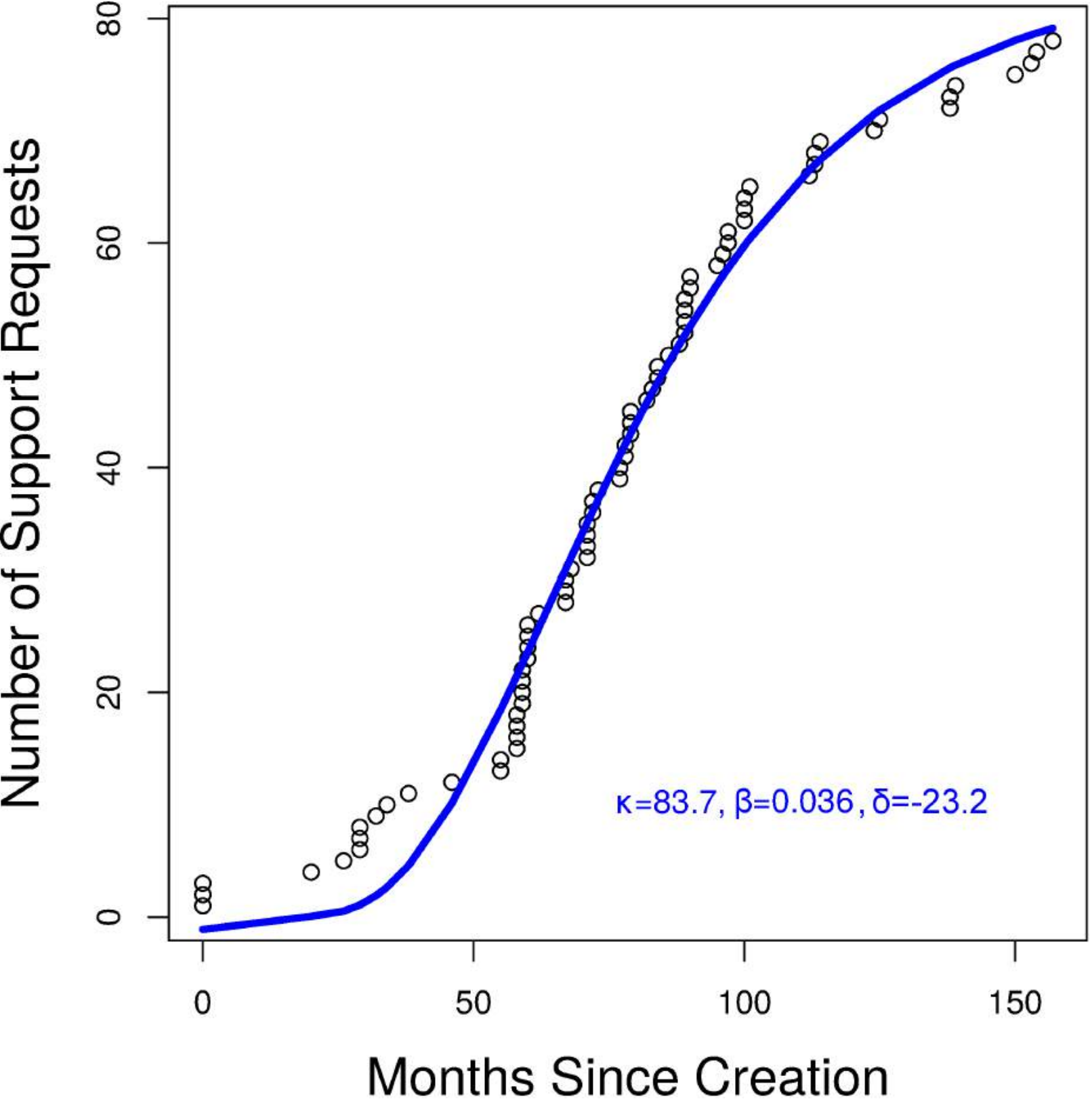}
\label{fig:example_fit}
\vspace{1mm}
\end{figure}

We applied LANLAN to model the rate of support requests and problem reports for multiple programs (in Eclipse and Bioconductor). Growth curves were trained on Q\&A forum data using non-linear least squares and Bayesian parameter estimation. Figure \ref{fig:example_fit} shows an example of fitting a curve to the growth of support requests for the ArrayExpress package in Bioconductor. In this case, the fitted parameters are 83.7 for $\kappa$, 0.036 for $\beta$ and -23.2 for $\delta$. The number of support requests grows exponentially at first, but then slows down, forming a classic S-curve shape. A possible explanation is that the way new software works may not be immediately clear so people submit lots of support requests, but as it develops to maturity, the documentation and interface improves, and users can look back at previous forum posts, so new support requests are not needed. Alternatively, the rate of new support requests may decrease as other packages become more popular, but ArrayExpress is still actively used (particularly with the rise of single cell analysis), so this seems less likely.

\begin{figure}[ht]
\caption{Comparing the Growth of Predicted Problem Reports in Eclipse from Q\&A Forum Data\\with Recorded Software Failures from Bug Reports}
\vspace{-4mm}
\centering
\includegraphics[width=.7\linewidth]{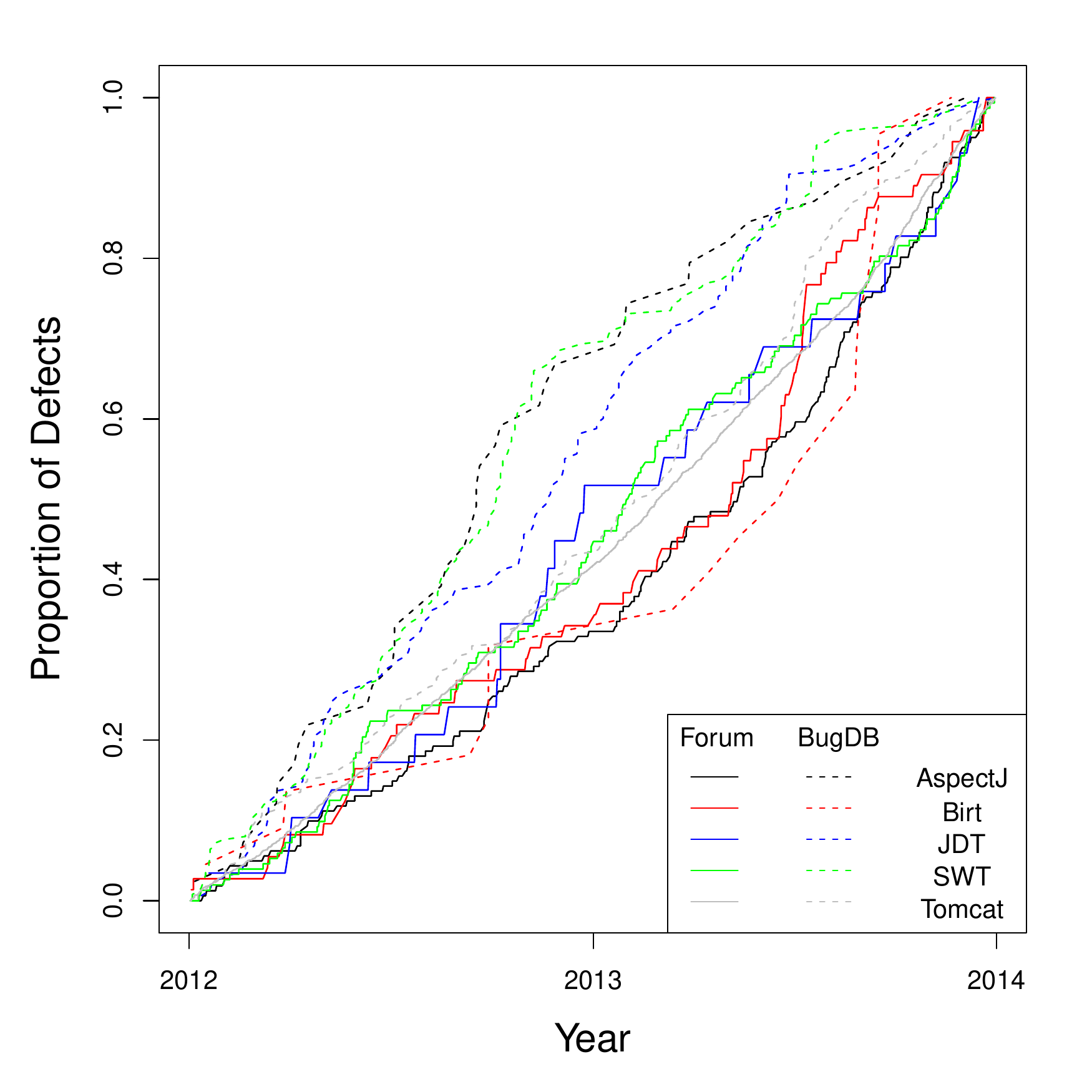}
\label{fig:growth_eclipse}
\end{figure}

We tested our predictions by training growth models on the occurrence of support requests and problem reports in the first $N$ months since creation, then evaluated their accuracy on the subsequent months. As more data (months) are added, the predictions move closer to the correct value. On average, we found that (when trained on the first half of the data), our prediction of the asymptote was only 6.2\% and 8.7\% away from the final value, for support requests and problem reports respectively. As a further independent evaluation, we compared our predictions for problem reports against the BugDB database for Eclipse \cite{ye_2014}. We observed some differences between the rate of problem reports in Q\&A forum data and software failures in BugDB (see Figure \ref{fig:growth_eclipse}), but the overall shape of the time series is similar (being suggestive of exponential growth) and pairwise tests of the area under the curve for each program showed no statistically significant differences (Student's t-test: p=0.152; Wilcoxon signed rank test: p=0.188). Neither the Q\&A forum nor bug tracking database provide complete information, and both are prone to random noise, but by combining them together we believe a more accurate estimation of problems encountered can be achieved.

\begin{table}[ht]
\vspace{4.5mm}
\centering
\begin{tabular}{cc|cc|cc}\toprule
& & \textbf{Mean}&\textbf{SD}&\textbf{W}&\textbf{P-value}\\\midrule
\multirow{2}{*}{$\kappa$} & \textbf{Problem Reports} & 32.4 &  34.9 & \multirow{2}{*}{33960} & \multirow{2}{*}{\num{1.17e-20}}\\
& \textbf{Support Requests} & 87.3 & 124\\\midrule
\multirow{2}{*}{$\beta$} & \textbf{Problem Reports} & 0.0414 &  0.0283 & \multirow{2}{*}{16785} & \multirow{2}{*}{\num{9.89e-05}}\\
& \textbf{Support Requests} & 0.0376 & 0.0365\\\midrule
\multirow{2}{*}{$\delta$} & \textbf{Problem Reports} & -22.0 &  15.9 & \multirow{2}{*}{19297} & \multirow{2}{*}{\num{0.0925}}\\
& \textbf{Support Requests} & -30.2 & 34.4\\
\end{tabular}
\caption{Mann-Whitney Tests for Curve Fitting}
\label{tab:wilcox_fitted_parameters}
\end{table}
\newpage

We compared growth curves fitted for problem reports against those for support requests. Comparing fitted parameters across all the Bioconductor packages, we found the upper asymptote ($\kappa$) to be significantly lower for problem reports than support requests, but the growth rate ($\beta$) was significantly higher (see Table \ref{tab:wilcox_fitted_parameters}). This suggests that, although the number of problem reports will ultimately be smaller, they grow more quickly; many programming issues are identified early, whereas support requests grow as the number of users increases. Although the average difference in the delay parameter ($\delta$) is small, the distributions differ considerably (symmetrical for problem reports, but skewed for support requests). Since the peak is further left for problem reports, the point of inflection for most packages will come earlier (if at all), which makes sense considering programming issues are often identified early. Nevertheless, the long tail to the left of the support requests' distribution means for some packages, the growth curve is monomolecular. These packages may be poorly written or documented (at least early in their life).

\subsection{Utility of Problem Reports and Support Requests (Answer to RQ4)}
To answer this question, we sampled 10 problem reports (Table \ref{tab:aspectj_problemreports}) and 10 support requests (Table \ref{tab:aspectj_supportrequests}) at random from AspectJ and investigated them to assess their potential relationship to software reuse. We also include the number of comments and answers, to give some indication of the response each question triggered. Comments in StackOverflow are generally used to make remarks and request further clarification, whereas answers are intended to provide solutions. We also indicate if code was included, and whether one of the answers was accepted by the original poster. In the case of problem reports, we also specify if a link is given at some point in the thread to a record of this problem in a bug database.

Interestingly, all but one of the problem reports (Table \ref{tab:aspectj_problemreports}) describe difficulties integrating AspectJ with other software. Spring (a framework providing aspect-oriented programming as part of its functionality) is frequently mentioned, along with Maven, IntelliJ, Kotlin, MinGW and Lombok. By contrast, only 2 out of the 10 support requests (Table \ref{tab:aspectj_supportrequests}) ask for help integrating AspectJ with other software (Maven and WebLogic). Although this is a small sample, it may suggest problem reports are more indicative of potential issues related to software reuse. The support requests were more often aimed at understanding a specific behaviour or functionality of the software. For example, one seeks to understand whether creating aspects will lead to the addition of new classes (actually, the changes are `weaved' into the existing classes) and another asks how to find the list of classes which meet particular pointcut criteria (pointcuts are join points at which AspectJ makes changes).

One support request\footnote{\url{https://stackoverflow.com/questions/4468097/why-pointcut-matchesstring-class-returns-true}} provided a code excerpt (see below) and asked why it was outputting ``true" when they expected it to output ``false". They had assumed that since String.class is not located inside the java.util package, it would not be found. However, the pointcut matches functions with the String.class type, so returns ``true". This behaviour is documented, but the user who submitted the question was new to AspectJ and became confused. Issues such as this may be ameliorated through detailed tutorials (or Vingettes in R), explaining how the software is expected to behave.

\begin{lstlisting}
public void test1() {
    AspectJExpressionPointcut pointcut = new AspectJExpressionPointcut();
    pointcut.setExpression("execution(public * java.util.*.*(..))");
    System.out.println(pointcut.matches(String.class));
}
\end{lstlisting}

The following code excerpt was taken from a problem report and it is intended to modify a function using two different advices (modifying functions), but the expresson did not match properly due to a potential bug (submitted to the eclipse bug database)\footnote{\url{https://stackoverflow.com/questions/41129938/aspectj-pointcut-matching-arguments-args-is-not-matching-correctly}}. A workaround is provided, using two different pointcuts, but it is unclear whether the bug has been fixed. Only two other problem reports in the 10 we sampled are associated with an entry in the bug database and these link to the same bug. As mentioned in the Background and Related Work section, Q\&A forum mining is essential because many bugs are not reported in bug databases, so cannot otherwise be taken into account for evaluation in software reuse.

\begin{lstlisting}
@Before("(execution(public static * business.security.service.LoginManagerHelper.authenticateUser(..)) && args( username, ..)) || "
        + "(execution(public static * webapp.util.LoginManagerAction.loginJAAS(..)) && args( *, *, username, ..))")
public void setUsername(JoinPoint jp, String username) {
    // inject the username into the MDC
    MDCUtils.setUsername(username);
}
\end{lstlisting}

Table \ref{tab:findings} provides a summary of some key findings from the study. It remains an open question whether projects with a large number of bug reports and support requests will be more difficult to reuse. Such an investigation is outside the scope of this paper and is challenging to answer without subjectivity. However, we have shown how problem reports and support requests encapsulate issues which could make software reuse more difficult, and in our sample, problem reports appear to be enriched in issues that occur when combining multiple software together. It is not necessary for reusers to train a prediction model on the software they are considering to reuse - they could instead make use of the features we have shown to be associated with problem reports and support requests; in future these may be included in a tool for software analysis (e.g. as an Eclipse plugin). For researchers interested in repeating our work to further investigate the issues concerning software reusability, we have found that a model trained on a small but diverse range of software is effective at predicting problem reports and support requests in other software.

\begin{table}[ht]
\vspace{4.5mm}
\centering
\begin{tabular}{c|c|c|c|c}\toprule
\multirow{2}{*}{\textbf{Paraphrased question}}&\textbf{\#Comments}&\textbf{Code}&\textbf{Accepted}&\textbf{Bug}\\
& \textbf{/Answers} & \textbf{Included?} &\textbf{Answer?} &  \textbf{DB?}\\\midrule
\textbf{AspectJ Maven plugin not executed} & 6/3 & Yes & Yes & No\\
\textbf{IntelliJ not working with AspectJ} & 4/1 & Yes & No & No \\
\textbf{AspectJ not working in Kotlin} & 12/4 & Yes & No & No \\
\textbf{Problem using AspectJ in MinGW} & 0/1 & Yes & Yes & No \\
\textbf{Difficulty configuring Spring security} & 1/1 & Yes & Yes & Yes \\
\textbf{Error when using Spring Roo} & 2/1 & No & No & No \\
\textbf{Spring security mode not working} & 1/2 & Yes & Yes & Yes \\
\textbf{Lombok not working with AspectJ} & 1/1 & Yes & No & No \\
\textbf{JUnit not working with Spring} & 4/0 & Yes & No & No \\
\textbf{Pointcut not matching correctly} & 5/1 & Yes & Yes & Yes \\
\end{tabular}
\caption{Problem Reports Sampled from AspectJ}
\label{tab:aspectj_problemreports}
\end{table}
%https://stackoverflow.com/questions/17916751/aspectj-maven-plugin-not-executed
%https://stackoverflow.com/questions/41475653/aspectj-not-working-using-intellij
%https://stackoverflow.com/questions/44364633/aspectj-doesnt-work-with-kotlin
%https://stackoverflow.com/questions/20393728/aspectj-cant-run-the-load-time-weaving-example
%https://stackoverflow.com/questions/28658718/spring-boot-spring-security-with-aspectj-not-working
%https://stackoverflow.com/questions/4952321/spring-roo-and-web-scaffolding-issue
%https://stackoverflow.com/questions/25207933/spring-security-aspectjmode-with-enableglobalmethodsecurity-not-working
%https://stackoverflow.com/questions/31360447/aspectj-gradle-lombok-does-not-work
%https://stackoverflow.com/questions/23518193/spring-aop-and-custom-annotations-not-working
%https://stackoverflow.com/questions/41129938/aspectj-pointcut-matching-arguments-args-is-not-matching-correctly

\begin{table}[ht]
\vspace{4.5mm}
\centering
\begin{tabular}{c|c|c|c}\toprule
\multirow{2}{*}{\textbf{Paraphrased question}}&\textbf{\#Comments}&\textbf{Code}&\textbf{Accepted}\\
& \textbf{/Answers} & \textbf{Included?} & \textbf{Answer?}\\\midrule
\textbf{Do aspects create new classes?} & 2/2 & No & Yes \\
\textbf{Sharing data with annotated method?} & 0/1 & No & No \\
\textbf{Why does my pointcut give this result?} & 0/1 & Yes & No \\
\textbf{Can exception handling be nested?} & 1/1 & Yes & Yes \\
\textbf{How to find methods from pointcut?} & 1/1 & Yes & No \\
\textbf{Static initialization checks in AspectJ?} & 1/2 & No & No \\
\textbf{How to cancel a method execution?} & 1/1 & Yes & Yes \\
\textbf{Help using AspectJ Maven plugin?} & 4/3 & Yes & Yes \\
\textbf{How to use AspectJ on WebLogic?} & 0/2 & Yes & Yes \\
\textbf{What is scattering and tangling?} & 0/1 & Yes & Yes \\
\end{tabular}
\caption{Support Requests Sampled from AspectJ}
\label{tab:aspectj_supportrequests}
\end{table}
%https://stackoverflow.com/questions/1560994/aspectj-problem
%https://stackoverflow.com/questions/46358538/share-some-variable-between-annotated-method-and-aspect
%https://stackoverflow.com/questions/4468097/why-pointcut-matchesstring-class-returns-true
%https://stackoverflow.com/questions/40656806/handling-a-specific-exception-type
%https://stackoverflow.com/questions/28452940/is-it-possible-to-return-value-from-spring-aspectj
%https://stackoverflow.com/questions/14518057/which-methods-are-being-covered-by-a-pointcut
%https://stackoverflow.com/questions/8053802/aspectj-add-static-initializer-to-class
%https://stackoverflow.com/questions/22507630/cancel-a-method-execution-in-an-aspect-which-has-been-catched-by-before
%https://stackoverflow.com/questions/23590662/aspectj-with-weblogic
%https://stackoverflow.com/questions/37547695/what-are-scattering-and-tangling-in-aop

\begin{table}[ht]
\centering
\begin{tabular}{C{0.3\textwidth}|C{0.3\textwidth}|C{0.3\textwidth}}
\toprule
\textbf{Finding} & \textbf{Implication} & \textbf{Relevant To}\\\midrule
\multirow{3}{0.3\textwidth}{$>$0.9 AUROC using default parameters} & Effective at distinguishing forum posts & Practitioners\\&&\\
 & Parameter tweaking may have limited impact, as performance already high & Researchers \\\midrule
\multirow{3}{0.3\textwidth}{Robustness improved by training on multiple diverse programs} & Should train on programs related to subject area & Practitioners \\&&\\
& Explore optimal training set for general use & Researchers \\\midrule
\multirow{3}{0.3\textwidth}{$>$95\% R$^2$ maintained using 5 features for problem reports / support requests} & The features selected can help indicate important aspects of reusability & Practitioners \\&&\\
& Possible to make effective predictions using reduced number of features & Researchers \\\midrule
R.Comments significant for support requests but not problem reports & Indicates the role documentation plays in reusability & Practitioners \\\midrule
Growth rate higher for problem reports than support requests & Risk of failure increases quickly if not addressed & Practitioners \\\midrule
\end{tabular}
\caption{Summary of some key findings and implications}
\label{tab:findings}
\end{table}

\section{Threats to Validity}
We addressed internal threats to validity using statistical metrics and tests: to evaluate the effectiveness of LANLAN for distinguishing problem reports and support requests, we used the area under the receiver operator curve (AUROC), precision and recall; when comparing their growth, we used Mann Whitney tests, reporting W as an effect size in Table \ref{tab:wilcox_fitted_parameters}; and when exploring the features that contribute to this growth (through association analysis), we applied Bonferroni correction to avoid p-values being significant by random chance. We also checked the assumptions of the models and tests we used (regarding the residuals and linearity) using plots produced by the stats package in R.

Experiments were repeated multiple times to improve the robustness of our results. For example, each Q\&A forum post was annotated three times and the label (problem report or support request) chosen by consensus. We evaluated the effectiveness of multiple classification algorithms, using the area under the receiver operator curve (AUROC), and compared our approach (LANLAN( to keyword-based alternatives. We trained and tested LANLAN on different programs and compared the growth of failures predicted by Q\&A forum posts (StackOverflow) and bug reports (BugDB). It is possible future researchers may be able to improve upon our results by changing some of the settings. For example, we used the default window size of 15 words in GloVe, as it has previously been shown to be effective, but adjusting this could increase the accuracy of prediction.

By focusing on the Eclipse and Bioconductor worked examples, there is an external threat to validity that LANLAN may not work on other software. Indeed, we have found our model to be sensitive to domain differences, since training it on one software (AspectJ) and applying it to others (JDT and edgeR) resulted in reductions in recall (although precision and AUROC were comparable to applying our model to the software on which it was trained). By training our model on multiple software, particularly those from different domains, we found it to be more robust when applying it to a new program (PROcess). However, a suitable question for future research would be whether this strategy would work if applying LANLAN to other domains. Eclipse is a widely used suite of programs for software development, being the subject of previous research into defect prediction, and Bioconductor is a large bioinformatics project containing diverse packages (from data processing to mathematical modelling and graphical interfaces). Since our worked examples represent prominent projects from two completely different fields, we believe they are likely to be representative of a wide range of other software.

\section{Conclusions}
We illustrated an approach (LANLAN) to classify Question and Answer (Q\&A) forum posts (into support requests and problem reports), such that they can be used to reveal information pertinent to reuse and reusability. We mined data from two large open source projects (Eclipse and Bioconductor), chosen for their differences in purpose as well as practices of reuse (systematic vs ad-hoc), increasing the likelihood LANLAN can be generalised to a wide range of software, particularly where more traditional resources (e.g. bug tracking databases) are unavailable. It is our belief that by integrating a greater variety of data and using sophisticated modelling techniques to analyse the results, the accuracy of the features identified and used for analysing software can be improved.

LANLAN achieved an AUROC of 0.930 in cross validation on a single program (AspectJ) and 0.970 AUROC when training the model on three programs and testing it on a fourth (PROcess). Growth curve analysis revealed the upper asymptote ($\kappa$) to be lower for problem reports (i.e. developers should expect more requests for clarification rather than issues with the code). However, the $\beta$ growth parameter was higher for problem reports, confirming software reusability issues related to defects increase more quickly at the beginning of the software's life. Cyclomatic complexity was more associated with support requests, whereas the issues identified by Codetools were more relevant to problem reports; complex software is more difficult to understand, but not necessarily more likely to be incorrect, whereas poor coding style often produces defects. These findings illustrate the effectiveness of LANLAN to classify Q\&A forum posts into useful categories (problem reports and support requests) for exploring potential software reusability metrics, revealing aspects of the various issues that can make software reuse more difficult. By improving understanding of the features that affect reusability, our research constitutes a first step towards the development of powerful new tools to assist software development. For example, the information gained from this study with regards to which metrics that are more indicative of problem reports or support requests could be used to automatically highlight potential reusability issues, and growth models can predict how quickly problems are likely to arise, thus guiding efficient management of focused interventions to improve reusability. Nevertheless, this endeavor would require considerable effort and may need to be tailored to different software fields.

\section{Code Availability}
The following code was used in this paper and is available from the links below:\\

\textbf{GloVe:} Word embedding

(\url{https://github.com/stanfordnlp/GloVe})\\

\textbf{MLR:} Machine learning 

(\url{https://github.com/mlr-org/mlr/})\\

\textbf{lm:} Association analysis

(\url{https://stat.ethz.ch/R-manual/R-devel/library/stats/html/lm.html})\\

\textbf{nls:} Least squares curve fitting

(\url{https://stat.ethz.ch/R-manual/R-devel/library/stats/html/nls.html})\\

\textbf{rjags:} Bayesian curve fitting 

(\url{https://cran.r-project.org/web/packages/rjags/index.html})

\bibliographystyle{elsarticle-num}

\bibliography{references}

\begin{thebibliography}{10}
\expandafter\ifx\csname url\endcsname\relax
  \def\url#1{\texttt{#1}}\fi
\expandafter\ifx\csname urlprefix\endcsname\relax\def\urlprefix{URL }\fi
\expandafter\ifx\csname href\endcsname\relax
  \def\href#1#2{#2} \def\path#1{#1}\fi

\bibitem{mohagheghi_2007}
P.~Mohagheghi, R.~Conradi, Quality, productivity and economic benefits of
  software reuse: a review of industrial studies, Empirical Software
  Engineering 12~(5) (2007) 471--516.

\bibitem{mcilroy_1969}
M.~McIlroy, Mass produced software components, in: P.~Naur, B.~Randell (Eds.),
  Nato science committee NATO, Scientific Affairs Division NATO, Belgium, 1969,
  pp. 1--136.

\bibitem{svahnberg_2016}
M.~Svahnberg, T.~Gorschek, A model for assessing and re‐assessing the value
  of software reuse, Journal of Software: Evolution and Process 29~(4) (2017)
  e1806.

\bibitem{frakes_2005}
W.~B. Frakes, K.~Kang, Software reuse research: Status and future, Transactions
  on Software Engineering 31~(7) (2005) 529--535.

\bibitem{ampatzoglou_2018}
A.~Ampatzoglou, S.~Bibi, A.~Chatzigeorgiou, P.~Avgeriou, I.~Stamelos,
  Reusability index: A measure for assessing software assets reusability, in:
  Proc. 17th Int. Conf. Software Reuse, Springer, 2018, pp. 43--58.

\bibitem{lemley_1997}
M.~Lemley, D.~O'Brien, Encouraging software reuse, Stanford Law Review 49~(2)
  (1997) 255--304.

\bibitem{endres_1975}
A.~Endres, An analysis of errors and their causes in system programs, IEEE
  Trans. Softw. Eng. SE-1~(2) (1975) 140--149.

\bibitem{goel_1979}
A.~L. Goel, K.~Okumoto, Time-dependent error-detection rate model for software
  reliability and other performance measures, IEEE Trans. Softw. Eng. R-28~(3)
  (1979) 206--211.

\bibitem{antoniol_2004}
G.~Antoniol, H.~Gall, M.~{Di Penta}, M.~Pinzger, {Mozilla: Closing the Circle},
  Tech. Rep. TUV-1841-2004-05, Technical University of Vienna (2004).

\bibitem{radjenovic_2013}
D.~Radjenovi\'{c}, M.~Heri\u{c}ko, R.~Torkar, A.~\u{Z}ivkovi\u{c}, Software
  fault prediction metrics: A systematic literature review, Information
  Software Technology 55~(8) (2013) 1397--1418.

\bibitem{bowes_2017}
D.~Bowes, T.~Hall, J.~Petri\'{c}, Software defect prediction: do different
  classifiers find the same defects?, Software Quality Journal (2017) 1--28.

\bibitem{hall_2012}
T.~Hall, S.~Beecham, D.~Bowes, D.~Gary, S.~Counsell, A systematic literature
  review on fault prediction performance in software engineering, IEEE Trans.
  Softw. Eng. 38~(6) (2012) 1276--1304.

\bibitem{jansen_2014}
S.~Jansen, Measuring the health of open source software ecosystems: Beyond the
  scope of project health, Information and Software Technology 56 (2014)
  1508--1519.

\bibitem{bedoya_2014}
O.~Franco-Bedoya, D.~Ameller, D.~Costal, X.~Franch, Queso: a quality model for
  open source software ecosystems, in: Proc. 9th Int. Conf. Software
  Technologies, IEEE, Washington, DC, 2014, pp. 39--62.

\bibitem{nguyen_2010}
T.~H.~D. Nguyen, B.~Adams, A.~E. Hassan, {A Case Study of Bias in Bug-Fix
  Datasets}, in: Working Conf. Reverse Engineering, 2010, pp. 259--268.

\bibitem{lotufo_2012}
R.~Lotufo, L.~Passos, C.~Krzysztof, {Towards improving bug tracking systems
  with game mechanisms}, in: Working Conf. Mining Software Repositories, 2012,
  pp. 2--11.

\bibitem{schugerl_2008}
P.~Schugerl, J.~Rilling, P.~Charland, {Mining Bug Repositories--A Quality
  Assessment}, in: Proc. Int. Conf. Computational Intelligence Modelling
  Control Automation, 2008, pp. 1105--1110.

\bibitem{sun_2011}
J.~Sun, {Why are Bug Reports Invalid?}, in: Proc. Int. Conf. Software Testing,
  Verification and Validation, 2011, pp. 407--410.

\bibitem{herzig_2013}
K.~Herzig, S.~Just, A.~Zeller, {It's not a bug, it's a feature: how
  misclassification impacts bug prediction}, in: Proc. Int. Conf. Software
  Engineering, 2013, pp. 392--401.

\bibitem{abdalkareem_2017}
R.~Abdalkareem, E.~Shihab, J.~Rilling, What do developers use the crowd for? a
  study using stack overflow, IEEE Software 34~(2) (2017) 53--60.

\bibitem{bachmann_2010}
A.~Bachmann, C.~Bird, F.~Rahman, P.~Devanbu, A.~Bernstein, {The Missing Links:
  Bugs and Bug-fix Commits}, in: Int. Symp Foundations Software Engineering,
  2010, pp. 97--106.

\bibitem{vasilescu_2013}
B.~Vasilescu, V.~Filkov, A.~Serebrenik, Stackoverflow and github: Associations
  between software development and crowdsourced knowledge, in: Proc. 6th Int.
  Conf. Social Computing, IEEE, 2013, pp. 1--54.

\bibitem{wang_2019}
D.~Wang, B.~K. Szymanski, T.~Abdelzaher, H.~Ji, L.~Kaplan, Software reuse
  research: Status and future, Computer 52~(1) (2019) 36--45.

\bibitem{whittaker_1998}
S.~Whittaker, L.~Terveen, W.~Hill, L.~Cherny, The dynamics of mass interaction,
  in: Proc. ACM conference on Computer supported cooperative work, ACM, New
  York, NY, 1998, pp. 257--264.

\bibitem{treude_2011}
C.~Treude, O.~Barzilay, M.-A. Storey, How do programmers ask and answer
  questions on the web?, in: Proc. 33rd Int. Conf. Software Engineering, ACM,
  New York, NY, 2011, pp. 804--807.

\bibitem{anderson_2012}
A.~Anderson, D.~Huttenlocher, J.~Kleinberg, J.~Leskovec, Discovering value from
  community activity on focused question answering sites: A case study of stack
  overflow, in: Proc. 18th ACM SIGKDD Int. Conf. Knowledge discovery and data
  mining, ACM, New York, NY, 2012, pp. 850--858.

\bibitem{nasehi_2012}
S.~M. Nasehi, J.~Sillito, F.~Maurer, C.~Burns, What makes a good code example?:
  A study of programming q\&a in stackoverflow, in: Proc. 28th Int. Conf.
  Software Maintenance, IEEE, Washington, DC, 2012, pp. 1063--6773.

\bibitem{yang_2011}
L.~Yang, S.~Bao, Q.~Lin, X.~Wu, Analyzing and predicting not-answered questions
  in community-based question answering services, in: Proc. 25th AAAI Conf.
  Artificial Intelligence, AAAI Press, Palo Alto, CA, 2011, pp. 1273--1278.

\bibitem{zhang_2015b}
Y.~Zhang, D.~Lo, X.~Xia, J.-L. Sun, Multi-factor duplicate question detection
  in stack overflow, J. Computer Science and Technology 30~(5) (2015)
  981–997.

\bibitem{ponzanelli_2014}
L.~Ponzanelli, G.~Bavota, M.~D. Penta, R.~Oliveto, M.~Lanza, Mining
  stackoverflow to turn the ide into a self-confident programming prompter, in:
  Proc. 11th Working Conf. Mining Software Repositories, 2014, pp. 102--111.

\bibitem{rong_2016}
X.~Rong, S.~Yan, S.~Oney, M.~Dontcheva, E.~Adar, Codemend: Assisting
  interactive programming with bimodal embedding, in: Proc. 29th User Interface
  and Software Technology Symposium, ACM, New York, NY, 2016, pp. 247--258.

\bibitem{kagdi_2007}
H.~Kagdi, M.~L. Collard, J.~I. Maletic, A survey and taxonomy of approaches for
  mining software repositories in the context of software evolution, J.
  Software: Evolution and Process 19 (2007) 77--131.

\bibitem{ray_2014}
B.~Ray, D.~Posnett, V.~Filkov, P.~Devanbu, A large scale study of programming
  languages and code quality in github, in: Proc. 22nd Int. Symp Foundations
  Software Engineering, ACM, New York, NY, 2014, pp. 155--165.

\bibitem{zanetti_2013}
M.~S. Zanetti, I.~Scholtes, C.~J. Tessone, F.~Schweitzer, Categorizing bugs
  with social networks: A case study on four open source software communities,
  in: Proc. 35th Int. Conf. Software Engineering, ACM, New York, NY, 2013, pp.
  1032--1041.

\bibitem{zhang_2015}
T.~Zhang, G.~Yang, B.~Lee, E.~K. Lua, A novel developer ranking algorithm for
  automatic bug triage using topic model and developer relations, in: Proc.
  21st Asia-Pacific Software Engineering Conference, IEEE, Washington, DC,
  2015, pp. 1530--1362.

\bibitem{manikas_2016}
K.~Manikas, Revisiting software ecosystems research: A longitudinal literature
  study, J. Systems and Software 117 (2016) 84--103.

\bibitem{mens_2014}
T.~Mens, M.~Claes, P.~Grosjean, Ecos: Ecological studies of open source
  software ecosystems, in: Proc. IEEE Conf. Software Maintenance,
  Reengineering, and Reverse Engineering, IEEE, Washington, DC, 2014, pp.
  403--406.

\bibitem{zeller_2013}
A.~Zeller, Can we trust software repositories?, in: J.~M\"{u}nch, K.~Schmid
  (Eds.), Perspectives on the Future of Software Engineering, Springer-Verlag,
  Heidelberg, 2013, pp. 209--2015.

\bibitem{bird_2009}
C.~Bird, A.~Bachmann, E.~Aune, J.~Duffy, A.~Bernstein, V.~Filkov, P.~Devanbu,
  Fair and balanced?: bias in bug-fix datasets, in: Proc. Int. Conf.
  Foundations Software Engineering, 2009.

\bibitem{goth_2016}
G.~Goth, Deep or shallow, nlp is breaking out, Commun. ACM 59~(3) (2016)
  13--16.

\bibitem{wittgenstein_1953}
Philosophical Investigations, Blackwell, Oxford, 1953.

\bibitem{firth_1957}
J.~R. Firth, A synopsis of linguistic theory, 1930-1955, Blackwell, Oxford,
  1957.

\bibitem{giatsoglou_2017}
M.~Giatsoglou, M.~G. Vozalis, K.~Diamantaras, A.~Vakali, G.~Sarigiannidis,
  K.~C. Chatzisavvas, Sentiment analysis leveraging emotions and word
  embeddings, Expert Systems With Applications 69 (2017) 214--224.

\bibitem{deerwester_1990}
S.~Deerwester, S.~Dumais, G.~Furnas, T.~Landauer, R.~Harshman, Indexing by
  latent semantic analysis, Expert Systems With Applications 41~(6) (1990)
  391--407.

\bibitem{mikolov_2013}
T.~Mikolov, I.~Sutskever, K.~Chen, G.~Corrado, J.~Dean, Distributed
  representations of words and phrases and their compositionality, in: Proc.
  26th Int. Conf. Neural Information Processing Systems, 2013, pp. 3111--3119.

\bibitem{nguyen_2017}
T.~V. Nguyen, A.~T. Nguyen, H.~D. Phan, T.~D. Nguyen, T.~N. Nguyen, Combining
  word2vec with revised vector space model for better code retrieval, in: Proc.
  Int. Conf. Software Engineering Companion, 2017.

\bibitem{pennington_2014}
J.~Pennington, R.~Socher, C.~D. Manning, Glove: Global vectors for word
  representation, in: Proc. Conf. Empirical Methods Natural Language
  Processing, 2014, pp. 1532--1543.

\bibitem{hendricks_2017}
L.~A. Hendricks, O.~Wang, E.~Shechtman, J.~Sivic, T.~Darrell, B.~Russell,
  Localizing moments in video with natural language, in: Int Conf. Computer
  Vision, 2017 [in press].

\bibitem{greenwald_2017}
A.~G. Greenwald, An {AI} stereotype catcher, Science 356~(6334) (2017)
  133--134.

\bibitem{rahman_2014}
F.~Rahman, S.~Khatri, E.~T. Barr, P.~Devanbu, Comparing static bug finders and
  statistical prediction, in: Proc. 36th Int. Conf. Software Engineering, ACM,
  New York, NY, 2014, pp. 424--434.

\bibitem{johnson_2013}
B.~Johnson, Y.~Soong, E.~Murphy-Hill, R.~Bowdidge, Why don't software
  developers use static analysis tools to find bugs?, in: Proc. 35th Int. Conf.
  Software Engineering, ACM, New York, NY, 2013, pp. 672--681.

\bibitem{ray_2016}
B.~Ray, V.~Hellendoorn, S.~Godhane, Z.~Tu, A.~Bacchelli, P.~Devanbu, On the
  ``naturalness" of buggy code, in: Proc. 38th Int. Conf. Software Engineering,
  ACM, New York, NY, 2016, pp. 428--439.

\bibitem{kenter_2015}
T.~Kenter, M.~{de Rijke}, Short text similarity with word embeddings, in: Proc.
  Int. Conf. Information and Knowledge Management, 2015.

\bibitem{bischl_2016}
B.~Bischl, M.~Lang, L.~Kotthoff, J.~Schiffner, J.~Richter, E.~Studerus,
  G.~Casalicchio, Z.~M. Jones, mlr: Machine learning in r, Machine Learning
  Research 17~(170) (2016) 1--5.

\bibitem{balding_2006}
D.~J. Balding, A tutorial on statistical methods for population association
  studies, Nature Reviews Genetics 7 (2006) 781--791.

\bibitem{dunn_1961}
O.~J. Dunn, Multiple comparisons among means, J. American Statistical
  Association 56~(293) (1961) 52--64.

\bibitem{panik_2014}
M.~J. Panik, Growth Curve Modeling: Theory and Applications, Wiley, Hoboken,
  NJ, 2014.

\bibitem{hummel_2008}
O.~Hummel, W.~Janjic, C.~Atkinson, Code conjurer: Pulling reusable software out
  of thin air, Software 25~(5) (2008) 45--52.

\bibitem{martinez_2016}
J.~Martinez, T.~Ziadi, M.~Papadakis, T.~F. Bissyand\'{e}, J.~Klein, Y.~L.
  Traon, Feature location benchmark for software families using eclipse
  community releases, in: Proc. 15th Int. Conf. Software Reuse, Springer, 2002,
  pp. 267--283.

\bibitem{martinez_2017}
J.~Martinez, T.~Ziadi, T.~F. Bissyand\'{e}, J.~Klein, Y.~L. Traon, Bottom-up
  technologies for reuse: automated extractive adoption of software product
  lines, in: Proc. 39th Int. Conf. Software Engineering, IEEE, 2017, pp.
  67--70.

\bibitem{ye_2014}
X.~Ye, R.~Bunescu, C.~Liu, Learning to rank relevant files for bug reports
  using domain knowledge, in: Proc. Int. Conf. Foundations Software
  Engineering, 2014.

\bibitem{gentleman_2005}
R.~C. Gentleman, V.~J. Carey, W.~Huber, R.~Irizarry, S.~Dudoit, Bioinformatics
  and Computational Biology Solutions Using R and Bioconductor, Springer, New
  York, NY, 2005.

\bibitem{brown_2002}
A.~W. Brown, G.~Booch, Reusing open-source software and practices: The impact
  of open-source on commercial vendors, in: Proc. 7th Int. Conf. Software
  Reuse, Springer, 2002, pp. 123--136.

\bibitem{chidamber_1994}
S.~R. Chidamber, C.~F. Kemerer, A metrics suite for object oriented design,
  IEEE Transactions on Software Engineering 20~(6) (1994) 476--493.

\bibitem{buse_2010}
R.~P.~L. Buse, W.~R. Weimer, Learning a metric for code readability, IEEE
  Transactions on Software Engineering 36~(4) (2010) 546--558.

\bibitem{mccabe_1976}
T.~J. McCabe, A complexity measure, IEEE Transactions on Software Engineering
  2~(4) (1976) 308--320.

\bibitem{ammann_2016}
P.~Ammann, J.~Offutt, Introduction to Software Testing, {Cambridge University
  Press}, Cambridge, United Kingdom, 2016.

\end{thebibliography}

\end{document}